**Partially Connected Automated Vehicle Cooperative Control Strategy with a Deep Reinforcement Learning Approach**

Haotian Shi
Graduate Research Assistant
[1]Department of Civil and Environmental Engineering
[1]University of Wisconsin, Madison
[2]School of Mechanical Engineering
[2]Tianjin University, Tianjin, China,
1208 Engineering Hall 1415 Engineering Drive
Madison, WI 53706
Email: hshi84@wisc.edu
Phone: (608) 571-8009

Yang Zhou (Corresponding Author)
Postdoctoral Researcher
Department of Civil and Environmental Engineering
University of Wisconsin, Madison
1208 Engineering Hall 1415 Engineering Drive
Madison, WI 53706
Email: zhou295@wisc.edu
Phone: (605) 300-0159

Keshu Wu
Graduate Research Assistant
Department of Civil and Environmental Engineering
University of Wisconsin, Madison
1208 Engineering Hall 1415 Engineering Drive
Madison, WI 53706
Email: kwu84@wisc.edu
Phone: (412)537-0930

Xin Wang
Assistant Professor
Department of Industrial System Engineering
3258 Mechanical Engineering Building 1513 University Ave
Madison, WI 53706
Email: xin.wang@wisc.edu
University of Wisconsin, Madison

Yangxin Lin
Graduate Research Assistant
School of Software and Microelectronic,
Peking University, Beijing, China.

Email: martinlin@pku.edu.cn
Phone: (86) 18811734550

Bin Ran
Professor
Department of Civil and Environmental Engineering
University of Wisconsin, Madison
1212 Engineering Hall 1415 Engineering Drive
Madison, WI 53706
Email: bran@wisc.edu
Phone: (608) 262-0052

**ABSTRACT**

This paper proposes a cooperative strategy of connected and automated vehicles (CAVs) longitudinal control for partially connected and automated traffic environment based on deep reinforcement learning (DRL) algorithm, which enhances the string stability of mixed traffic, car following efficiency, and energy efficiency. Since the sequences of mixed traffic are combinatory, to reduce the training dimension and alleviate communication burdens, we decomposed mixed traffic into multiple subsystems where each subsystem is comprised of human-driven vehicles (HDV) followed by cooperative CAVs. Based on that, a cooperative CAV control strategy is developed based on a deep reinforcement learning algorithm, enabling CAVs to learn the leading HDV's characteristics and make longitudinal control decisions cooperatively to improve the performance of each subsystem locally and consequently enhance performance for the whole mixed traffic flow. For training, a distributed proximal policy optimization is applied to ensure the training convergence of the proposed DRL. To verify the effectiveness of the proposed method, simulated experiments are conducted, which shows the performance of our proposed model has a great generalization capability of dampening oscillations, fulfilling the car following and energy-saving tasks efficiently under different penetration rates and various leading HDVs behaviors.

Keywords: partially connected automated traffic environment, cooperative control, deep reinforcement learning, traffic oscillation dampening, energy efficiency

## 1. INTRODUCTION

With the fast development of vehicular automation and communication technology, connected and automated vehicles (CAV) gradually occupied some portion of the vehicle market. In the very near future, CAVs and human-driven vehicles (HDV) are envisioned to be driven on the same road or even the same lane. This transforms the traffic environment from pure human-driven vehicles to a partially connected and automated one, widely known as the mixed traffic environment (Zhou et al., 2019). Despite the traffic environment changes, traffic oscillations (i.e., the stop-and-go phenomenon) still remain challenging for mixed traffic. Small traffic oscillations can be amplified and result in detrimental impacts on various aspects of the transportation system (e.g., capacity, safety, energy consumption, and greenhouse gas emissions). Owing to CAV's communication and automation capability, such as adaptive cruise control (ACC) (e.g., Bageshwar et al., 2004, Takahama & Akasaka, 2018), cooperative adaptive cruise control (CACC) (e.g., Swaroop & Hedrick, 1996, Knorn et al., 2014), advanced vehicle control strategies are developed to enhance traffic flow performance in terms of traffic oscillation dampening (e.g., Cui et al., 2017, Stern et al., 2018, L. Li & Li, 2019), car following efficiency (Zhou, Wang, et al., 2019), and driving comfort elevating.

The study of vehicle control strategies attracted many researchers, though the majority of them only focused on the pure connected automated environments (Gong & Du, 2018). The mainstream methods can be largely divided into three categories based on their modeling differences: (i) linear or non-linear CAV longitudinal controller, (ii) model predictive control (MPC) based CAV longitudinal controller with functions of objectives and constraints, (iii) reinforcement learning (RL) based CAV longitudinal controller. First, linear (e.g., Stipanović et al., 2004; Naus et al., 2010; Morbidi et al., 2013) and non-linear (e.g., Bando et al., 1995; Treiber et al., 2000) CAV longitudinal controllers have closed-form formulations with parameters or gains. This type of model requires less calculation time due to its simplicity. In addition, its stability analysis is convenient due to the closed-form representation (e.g., Shladover et al., 2015; Petrillo et al., 2018; Yu Wang et al., 2020), and theoretically, the local stability and string stability of controllers can be guaranteed through appropriate parameter tuning. However, these linear and non-linear controllers have difficulties in designing an explicitly formulated framework to incorporate multiple control objectives (e.g., car following efficiency, string stability, energy efficiency) and collision-free constraints within reasonable vehicle acceleration/deceleration boundaries. Considering this limitation, MPC based CAV longitudinal controllers (e.g., Wang et al., 2016; Gong et al., 2016; Zhou et al., 2017; Gong & Du, 2018; Zhou et al., 2019;) have been favored in recent years. As an optimal control method, the MPC-based controller provides a flexible, constrained optimization framework that incorporates flexible optimizing objectives and constraints. Thus, CAV longitudinal control problem that considers string stability, efficiency, fuel consumption, and driving comfort can be solved at each timestep under safety constraints within a bounded acceleration range. Besides, this approach can provide driving decisions by predicting the future state of the leading vehicle trajectory, thus improving optimization and control performance in a rolling/receding horizon fashion. However, this approach is normally computational demanding, which is not applicable for real-time implementation (Zhou et al., 2017). Furthermore, it is challenging to quantitatively guarantee its platoon string stability due to the formulation complexity (Zhou et al., 2019).

Similar to the MPC controller, the recent breakthrough in the RL community provides alternative algorithms to be utilized (Karnchanachari et al., 2020). The advantages of RL based approaches are mainly reflected in the two aspects. First, an RL is a model-free and learning-based method, which is suitable for capturing complex system characteristics. Second, the computational burden of a RL algorithm mainly lies in its offline training process, while the learned driving policy can be rapidly implemented in real-time (Görges, 2017). Recently, RL algorithms have been gradually applied to design CAV controllers (Chong et al., 2013; Li et al., 2020; Zhou et al., 2020). Guan et al. (2019) applied an RL algorithm to cooperatively control CAVs at intersections and address the computational burden by training offline. Duan et al. (2020) comprehensively consider both high-level and low-level motion control for CAVs based on RL, which achieves a smooth and safe decision-making process. Wang et al. (2019) developed a Q-learning based bird-view approach for CAV control, which shows great control performance under complicated traffic environments. However, these works have not considered the mixed platoon's string stability and mainly focus on distance tracking (e.g., maintaining a car following headway with reasonable acceleration control). As far as the authors know, only Qu et al. (2020) proposed a control strategy based on the DDPG algorithm to dampen traffic oscillations and improve energy efficiency. However, this paper only considers non-cooperative vehicle control without exploiting information sharing. Besides, this study adopts a model-free gap policy, which cannot directly guarantee a stable traffic flow. In general, RL algorithm applications are still rare from the perspective of stability analysis in a mixed traffic environment.

Although there have been many studies about different approaches to optimizing car-following behavior and traffic flow, gaps in the studies on partially connected automated traffic environment remain in the following two aspects. Firstly, a fast computing cooperative multi-objective CAVs control algorithm for the mixed platoon to improve mixed traffic string stability (e.g., Naus et al., 2010; Ge & Orosz, 2014), car following efficiency, and energy efficiency, is still challenging. An exact optimization-based control such as MPC in a partially connected automated traffic environment is hard to construct due to the uncertainty of HDVs (Zhou et al., 2019), and fast computation is required to satisfy the real-time implementation. Secondly, it is hard to optimize mixed traffic flow considering the different combinations of CAVs and HDVs. Mixed platoons with different combinations have diverse characteristics, making it hard to develop a comprehensive model for optimal control, especially with multiple objectives.

Considering the pros and cons of different CAV longitudinal controllers and the gap in mixed-traffic studies, this research aims to develop a CAV longitudinal control strategy for a partially connected automated traffic environment from the following aspects. (i) Integrate the merits of reinforcement learning (RL) algorithm (suitable for complex characteristics and rapid online implementation) and MPC control methods (constrained optimization framework which enables multiple objectives and constraints based on the concept of equilibrium point). (ii) Achieve partially cooperative CAV longitudinal control with incorporating characteristics of HDVs, enhancing the whole mixed traffic flow while reducing the computation burdens by locally improving the traffic performance. (iii) Consider the sequence and combination of CAV and HDV and its impact on mixed traffic flow.

Specifically, we develop a cooperative CAV longitudinal control strategy for a partially connected automated traffic environment based on reinforcement learning. This study first decomposes a mixed platoon into multiple subsystems, consisting of one leading HDV followed by multiple cooperative CAVs to overcome the training difficulties and dimensions, reduce the computational cost, and alleviate communication burdens. Thus, different subsystems can comprise a mixed platoon of any CAV-HDV topology. The longitudinal acceleration of CAVs within each subsystem is cooperatively controlled by a well-designed RL agent. Field collected HDV trajectory data are used in the training process to guarantee the fidelity of HDV behaviors. Further, to achieve a desired control objective (incorporating the string stability performance of the mixed traffic, car following efficiency, and energy consumption), the reward function of RL agent is carefully built.

We select the Distributed Proximal Policy Optimization (DPPO) (Heess et al., 2017) algorithm, one of the most effective RL algorithms that enable continuous action, to develop the RL agent. DPPO algorithm derives from Trust Region Policy Optimization (TRPO) (Schulman et al., 2015) algorithm, which improves the convergence of policy updates by restricting Kullback-Leibler (KL) divergence between the prediction distribution of the old strategy and new strategy on the same batch of data. Then, the Proximal Policy Optimization (PPO) algorithm (Schulman et al., 2017), the default RL algorithm of Open AI, is similar to TRPO but shows better sample efficiency due to multiple updates per batch sample. Combined with the advantage of the Asynchronous Advantage Actor-Critic (A3C) algorithm, Google DeepMind proposed a distributed PPO (DPPO) algorithm to update the policy of the global agent in parallel through multiple working agents (Heess et al., 2017), which greatly improves training efficiency. Thus, this algorithm is adopted due to its support for continuous action and great convergence performance.

The contribution of the paper is mainly three-fold. First, we inherit the merits of reinforcement learning algorithm and MPC control methods by designing a novel reward function. Thus, the DRL algorithm framework can incorporate the constrained optimization framework merits based on the concept of equilibrium point to achieve a better control performance, especially for string stability performance. Secondly, the proposed control strategy decomposed mixed platoon and CAVs are controlled cooperatively and locally with incorporated HDV characteristics. In this way, CAVs based on the proposed deep reinforcement learning algorithm can make the longitudinal control decisions cooperatively by learning characteristics of the first leading HDV to improve each subsystem's performance locally and consequently enhance performance for the whole mixed traffic flow. Thirdly, the impact of the combination of CAVs and HDVs in the partially connected and automated environment is studied. This will provide guidance and reference for future research that considers lane-changing maneuvers.

This paper is organized as follows. Section 2 provides the mixed traffic setting. Then the model development of vehicle control strategies is in Section 3. In particular, Section 3.1 describes the proposed control strategy scheme in a Markov Decision Process (MDP), which builds the basis for constructing an RL algorithm. Based on the scheme, Section 3.2 and Section 3.3 provide the details of model development, consisting of RL algorithm implementation details by utilizing DPPO and training procedures. Section 4 analyzes the results of simulated experiments for model validation and application. The final section concludes the work.

## 2. ENVIRONMENT SETTING

This section describes the longitudinal control of CAVs in a partially connected automated traffic environment on a one-lane highway segment. For this research, we consider a single-lane straight highway segment of infinite length, and the lateral vehicle movement is neglected in this study. Besides, assumptions for the communication settings are made as: (i) Vehicle state information (e.g., speed, position, gap) can be shared and obtained by local CAVs through V2X communication in real-time. (ii) The communication delay of V2X is negligible in this study by the increasing maturity of 5G communication technology. (iii) Topology information of the mixed platoon can be obtained and delivered to CAVs by a roadside infrastructure.

For a mixed vehicular platoon consisting of N vehicles where N $\in Z^{+}$, we have a platoon topology:

$$\overrightarrow{T_p} = \{\gamma_1, \gamma_2, \gamma_{3,} \dots . \gamma_N\} \quad (1)$$

where $\gamma_i = \begin{cases} 0, \text{ if vehicle i is a CAV} \\ 1, \text{ if vehicle i is a HDV} \end{cases}$

There are $2^{N}$ combinations of $\gamma_i$ for a given $\vec{T}_P$, as shown in Figure 1. Different combinations and compositions diversify the mixed traffic environment, making it hard to develop a comprehensive model for real-time optimal control, especially the model with multiple objectives (e.g., car following efficiency, string stability, energy efficiency).

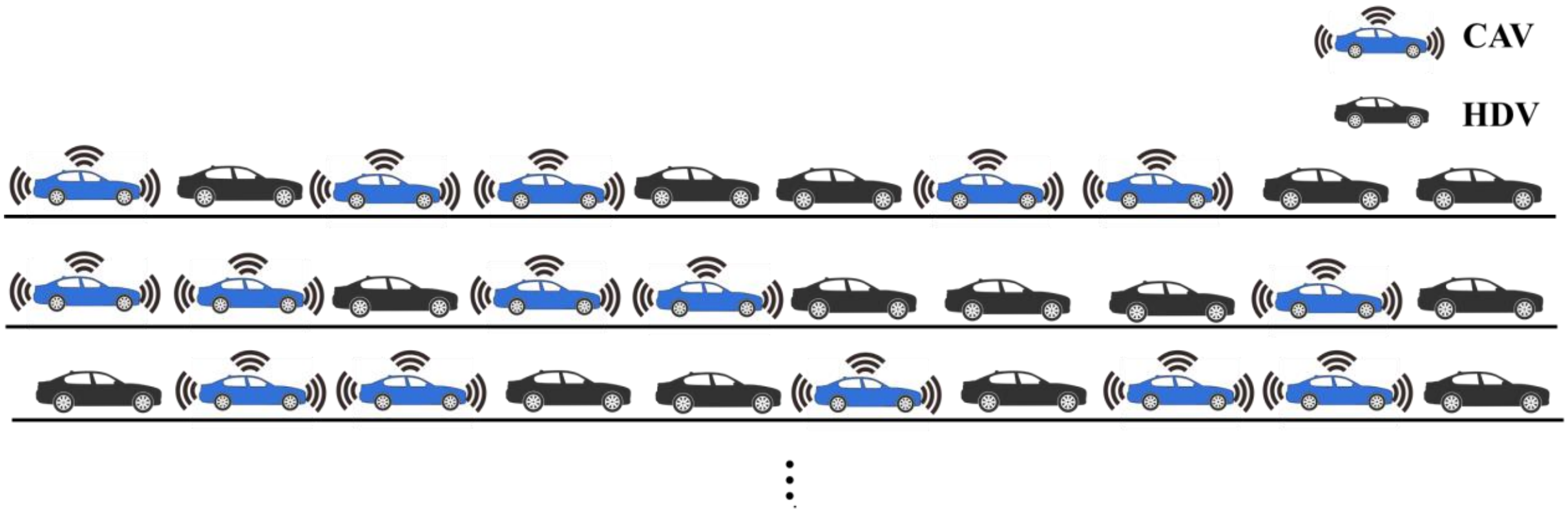

**Fig. 1.** Different compositions of partially connected automated traffic environment

To reduce the dimensions of the problem, we divide a mixed platoon into $N$ categories of the subsystem, each of which consists of one leading HDV followed by multiple CAVs. A subsystem with $k$ following CAVs is denoted by Subsystem $k$, $k = 1,2, \dots, N$, as seen in Figure 2 (Wang, 2018). With V2X communication and multiple sensors equipped, controlled CAVs within a certain distance behave cooperatively based on the shared state information (e.g., speed, position, gap). To this end, for each Subsystem $k$, a corresponding integrated Control Module for $k$ CAVs, denoted by $\mathrm{M}_k$, is developed to achieve their cooperative longitudinal control.

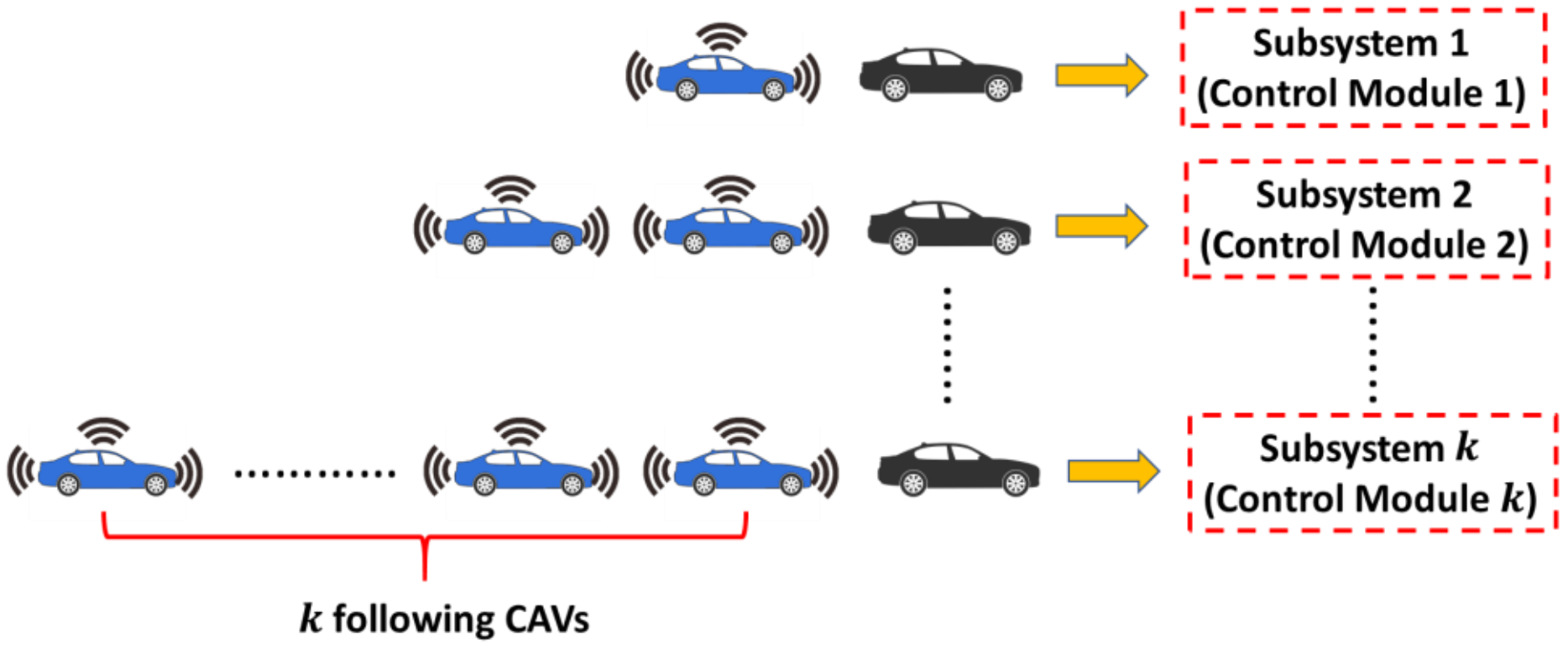

**Fig. 2.** Categories of subsystem

$N$ types of subsystems with $N$ control modules can comprise a mixed platoon of any order of HDVs and CAVs. Note that the specific HDV-CAV topology information of a platoon is assumed to be known in advance from the roadside infrastructure. There are three scenarios for a generally mixed platoon for subsystem decomposition, as shown in Figure 3. For scenario 1, if there is only one HDV between multiple CAVs, the HDV will be regarded as the leading vehicle (leader 2) of the adjacent module. For scenario 2, if a consecutive number of CAVs exceeds the control module size $N$ (e.g., $N = 5$), the $N$th CAV will be regarded as the leading vehicle of its adjacent upstream module. Thus, CAVs from 1 to 5 are controlled by $\mathrm{M}_5$ while CAV 6 and CAV 7 are controlled by $\mathrm{M}_2$. For scenario 3, if there are multiple HDVs between two CAVs, the last HDV will be set as the leader of the next subsystem. One can easily verify the three scenarios cover all topology information.

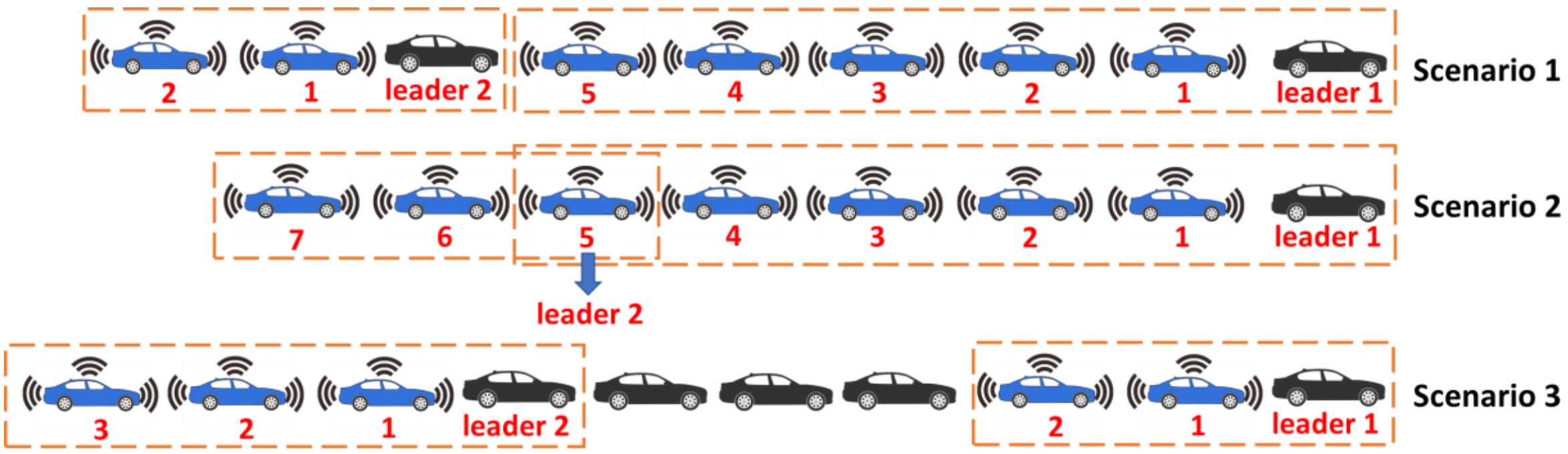

**Fig. 3.** Three scenarios for subsystem connection

Therefore, for a given mixed platoon, it is firstly decomposed into multiple subsystems based on the topology information, where CAVs within each subsystem are assigned to the corresponding control module. Then for the control module $k$, the DRL-based model $\mathrm{M}_k$ outputs acceleration signals to implement longitudinal control for CAVs based on the received shared state information within the subsystem at each timestep. Thus, the traffic flow in the partially connected automated traffic environment is optimized locally and separately. In addition, each module is separated by HDV so that string stability can be even strictly guaranteed within the subsystem, which simplified the stability challenge.

## 3. MODEL DEVELOPMENT

This section develops the controller model. The control scheme and formulation, the adopted RL algorithm, and training procedure are discussed. To clarify the methodology, car following state ***S***, system action ***A***, RL agent policy, and the design of reward ***R*** are defined in Section 3.1. Based on that, the RL algorithm and training procedure are given in Section 3.2 and Section 3.3.

### *3.1. Control Scheme and Formulation*

This section describes the design and formulation of model development. The modeling basis for reinforcement learning is the Markov Decision Process (MDP) (Van Otterlo & Wiering, 2012), which contains a set of two interactive objects: RL agent (CAV control algorithm) and environment (simulation platform). The two objects consist of state, action, policy, and reward (***S, A,π, R***). The interaction loop is illustrated in Figure 4. Through continuous interactions between the RL agent and the environment, the RL agent updates policy to adjust the output action ***A*** based on state ***S*** and reward ***R*** obtained from the environment and finally achieves the optimal policy. The detailed description of the four elements are described below:

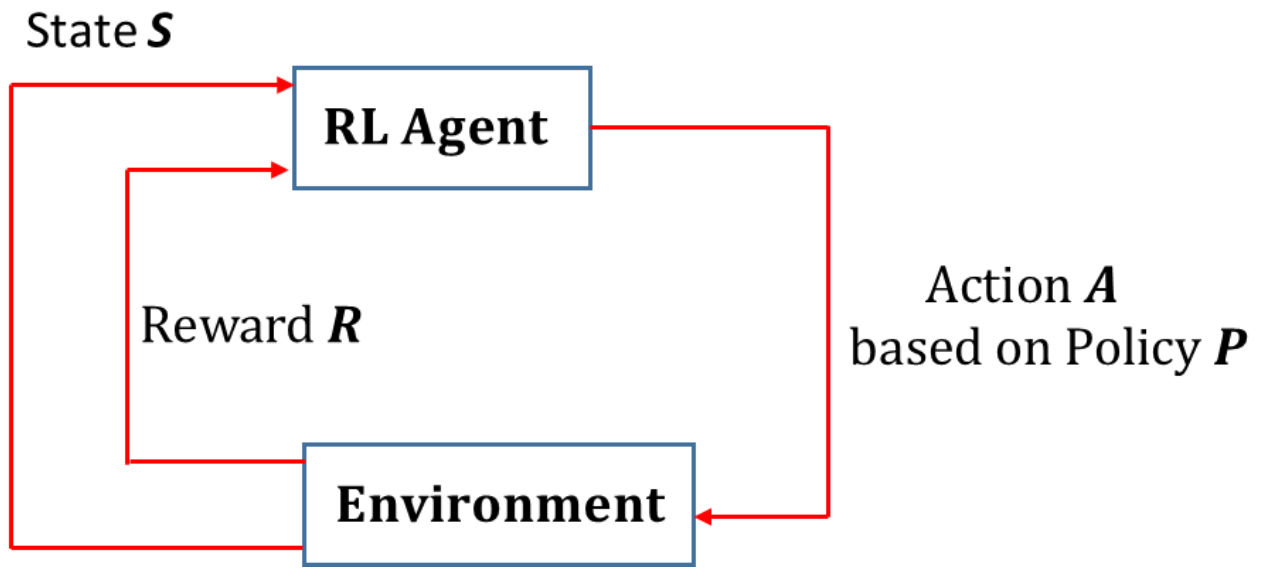

**Fig. 4.** Basic interaction loop of reinforcement learning algorithm

**Car Following State representation**

For the car following task, the state is normally defined as the instantaneous state information of the vehicle or vehicular platoons such as speed, space gap, and speed difference with the leading vehicle in DRL-based CF researches (Zhu et al., 2018; Qu et al., 2020), without a concept of equilibrium. These model-free gap policies allow the arbitrary change of the gap, making the traffic flow unstable. To remedy the issue, in this study, we incorporate the 'equilibrium' concept in the control theory field and define the vehicles' CF equilibrium spacing, which follows the constant time gap policy according to the Society of Automotive Engineer Standard (SAE). Specifically, the policy regulates CAVs to maintain the preset equilibrium spacing, which is proportional to the following vehicle speed. It regulates that the following vehicles have the same speed as their corresponding preceding vehicles. Thus, we integrate this CF strategy into the DRL algorithm framework defined as:

$$d_i^*(t) = v_i(t) * \tau_i^*(t) + l_i \tag{2}$$

where $d_i^*(t)$ is the target equilibrium spacing of vehicle at timestep t; $v_i(t)$ is the speed of vehicle $i$ at timestep t; $\tau_i^*(t)$ is the constant or time varying headway, and $l_i$ is the standstill spacing of vehicle $i$. Thus, the deviation from equilibrium spacing $\Delta d_i(t)$ and the relative speed with preceding vehicle $\Delta v_i(t)$ are specified as:

$$\Delta d_i(t) = d_i(t) - d_i^*(t) \tag{3}$$

$$\Delta v_i(t) = v_{i-1}(t) - v_i(t) \tag{4}$$

To incorporate the CF strategy, $\Delta v_i(t)$ and $\Delta d_i(t)$ are embedded in the vehicle state information. Specifically, $S_k^t$ is used to denote the immediate state of module *k* at time step t, which consists of immediate states for all following CAVs. Thus, the local vehicle information can be used for control at each timestep. $S_k^t$ is defined as:

$$S_k^t = \left[s_{k,1}^t \;\; s_{k,2}^t \;\; \dots \;\; s_{k,i}^t\right]^T,\; i \in [1,2 \dots k] \tag{5}$$

where each element $s_{k,i}^t = [v_i(t) \;\; g_i(t) \;\; \Delta v_i(t) \;\; \Delta d_i(t)]^T$ indicates the immediate state for CAV $i$ ($i^{th}$ following CAV) at timestep t, which includes four components: speed $v_i(t)$, inter-vehicle gap $g_i(t)$, the relative speed with the preceding vehicle $\Delta v_i(t)$, and deviation from equilibrium spacing $\Delta d_i(t)$, where the latter two terms we aim to regulate them to zero.

**Action representation**

System action ***A*** is output by RL agent to control all CAVs within a control module cooperatively. Similar to the state representation, $A_k^t$ denotes immediate acceleration sets for all following CAVs within the control module $k$, defined as:

$$A_k^t = \left[a_{k,1}^t \;\; a_{k,2}^t \dots a_{k,i}^t\right]^T \;, i \in [1,2 \dots k] \tag{6}$$

where each element $a_{k,i}^t$ indicates the immediate longitudinal acceleration to control CAV $i$ at timestep t.

**Policy representation**

Policy $\pi(a|s)$ is an implicit function trained by RL, which used to determine system action ***A*** based on current system state ***S***. Specifically, it decides the probability of taking a possible action $a$ on the state $s$ in the system, expressed by Eq. (7):

$$\pi(a|s) = \pi[A_k^t = a | S_k^t = s] \tag{7}$$

**Reward design**

In RL, the reward function is used to describe the control targets. A proper design of reward can significantly improve CAV CF behaviors. Three objectives, including car following efficiency, string stability fuel consumption, are considered from a traffic and energy efficiency perspective.

The car following efficiency is to measure the local stability for each car following pair around the equilibrium state, which is represented by quadratic forms of deviation from equilibrium state as suggested by Zhou et al., (2019):

$$c_i^t\left(x_i^t, a_i^t\right) = \; (x_i^t)^T Q_i x_i^t \tag{8}$$

where $x_i^k \; = \; [\Delta d_i(t),\; \Delta v_i(t)]$; $Q_i$ is a positive definite diagonal coefficients matrix, designed as the diagonal matrix below:

$$Q_i = \begin{bmatrix} \alpha_{1,i} & \\ & \alpha_{2,i} \end{bmatrix},\; \alpha_{1,i},\; \alpha_{2,i} \; > 0 \tag{9}$$

Further, a trade-off considering driving comfort in car-following control is also incorporated in the running cost $l_i\left(x_i^t, a_i^t\right)$ formulated below (Zhou et al., 2019):

$$l_i\left(x_i^t, a_i^t\right) = \; c_i^t\left(x_i^t, a_i^t\right) + \; M_i(a_i^t)^2 \tag{10}$$

where $(x_i^t)^T Q_i x_i^t$ is the control efficiency cost; $M_i(a_i^t)^2$ is defined as the comfort cost that evaluates driving comfort (Wang et al., 2014).

String stability is to measures the performance of the proposed control strategy in dampening disturbances through a platoon in the control module. The cumulative dampening ratio of the following CAV $i$ $d_{p,i}$ (Zhou et al., 2019), is defined to evaluate the string stability:

$$d_{p,i} = \frac{\|a_i(s)\|_2}{\|a_0(s)\|_2} \tag{11}$$

where $i$ is the index of the vehicle. Index 0 indicates the leading HDV of a control module. The control strategy aims to ensure that every CAV within the module $k$ is head-to-tail string stable, which satisfies:

$$d_{p,i} \leq 1, \forall i \in [1, 2 \ldots k] \tag{12}$$

To evaluate energy efficiency in a mixed platoon, we use the VT-micro model to quantify instantaneous fuel consumption. VT-Micro model, developed by (Rakha & Ahn, 2003), is a typical vehicle-based fuel consumption model related to vehicle speed and acceleration, which is shown in Eq. (13). The separated acceleration and deceleration show the precise predicting performance of real-world fuel consumption. Parameters used by (Ma et al., 2017) are adopted in this paper.

$$e_i(v_i^t, a_i^t) = \exp[\sum_{i=0}^{3}\sum_{j=0}^{3} K_{ij}(a_i^t)\left(\lfloor v_i^t \rceil_0^{v_f}\right)^i \left(\lfloor a_i^t \rceil_{a_{min}/sec}^{a_{max}/sec}\right)^j] \tag{13}$$

where $e_i(v_i^t, a_i^t)$ is the instantaneous fuel consumption rate $(ml/s)$; $K_{ij}$ is the polynomial coefficient; $v_f$ is the free flow speed.

Based on the discussed three objectives, an infinite-horizon optimal control problem on the basis of reward function is formulated. For module $k$ at timestep $t$, the optimal policy $\pi^*$ aims to maximize the discounted cumulative rewards in the infinite time horizon:

$$\pi^* = \arg\max_{\pi} \sum_{m=0}^{\infty} \Upsilon^m R(S_k^{t+m}, A_k^{t+m}) \tag{14}$$

where $R(S_k^t, A_k^t)$ represents the reward function, which indicates the immediate reward for module $k$ at timestep $t$. Specifically, it is the sum of immediate rewards of all CAVs within module $k$, defined as:

$$R(S_k^t, A_k^t) = \sum_{i=1}^{k}\left[r_{k,i}^t\left(s_{k,i}^t, a_{k,i}^t\right)\right] \tag{15}$$

where $r_{k,i}^t\left(s_{k,i}^t, a_{k,i}^t\right)$ is the immediate reward value of CAV $i$ at timestep $t$. To solve the optimal control problem, we incorporate the three objectives and soft constraints of traffic safety into the immediate reward $r_{k,i}^t$, which consists of two aspects: the original reward $r_{i,o}^t$ and the penalty reward $r_{i,p}^t$:

$$r_{k,i}^t = r_{i,o}^t + r_{i,p}^t \tag{16}$$

Specifically, the original reward $r_{i,o}^t$ considers both the car following efficiency and energy efficiency with adjustable weights to balance the control performance, which is formulated below:

$$r_{i,o}^t = \exp\left[-\left(C_i * l_i\left(x_i^t, a_i^t\right) + F_i * \mathrm{e}\left(v_i^t, a_i^t\right)\right)\right] \tag{17}$$

where $C_i$ and $F_i$ are the reward coefficients of the car following efficiency and energy efficiency, respectively. The penalty reward $r_{i,p}^t$ is to penalize the violation of string stability, safety time gap, and speed limit, which is shown in the following equation:

$$r_{i,p}^t = r_{i,pstablity}^t + r_{i,pgap}^t + r_{i,pspeed}^t \tag{18}$$

$$r_{i,pstablity}^t = \begin{cases} S_i, & \text{when dampening ratio } d_{p,i}(t) > 1 \text{ for every } n_d \text{ timesteps} \\ 0, & \text{when dampening ratio } d_{p,i}(t) \leq 1 \text{ for every } n_d \text{ timesteps} \end{cases} \tag{19-a}$$

$$r_{i,pgap}^{t} = \begin{cases} G_i, & \text{when time gap } \frac{g_i(t)}{v_i(t)} < g_t \\ 0, & \text{when time gap } \frac{g_i(t)}{v_i(t)} \geq g_t \end{cases}$$

(19-b)

$$r_{i,pspeed}^{t} = \begin{cases} D_i, & \text{when } v_i > v_f \\ 0, & when \ v_i \leq v_f \end{cases} \quad (19\text{-c})$$

where $S_i$, $G_i$, and $D_i$ are the negative penalty reward parameters for string stability, time gap, and speed. The threshold $g_t$ referred in (Milanés & Shladover, 2014) is adopted for the safety time gap.

It is worth noting that the optimal control problem defined in this study can also be formulated in the constrained optimization framework where the reward function (15) can be formulated as the stage cost; the soft penalty reward (18) can be converted to the hard constraints; the objective function (14) can be redefined to minimize the stage cost of each timestep with the finite time horizon. However, it is very challenging to maintain string stability and achieve real-time computation through constrained optimization due to the complex formulation, which is the reason and the motivation the DRL-based control strategy is adopted to solve this problem. Based on the approach scheme, we formulate a multi-objective optimal control problem and solve it in an RL context by developing corresponding DRL based models offline, leading to a DRL-based fast computing cooperative multi-objective CAVs control algorithm for a partially connected automated traffic environment.

### *3.2. Distributed Proximal Policy Optimization (DPPO) Algorithm*

For reinforcement learning (RL), model training aims to update the RL agent's policy to achieve an optimal solution. This section presents the principle of the Distributed Proximal Policy Optimization (DPPO) algorithm for policy updating (Heess et al., 2017). DPPO algorithm has an actor-critic structure with parameters updated both in the actor and critic network through the training process. Figure 5 demonstrates the principle of the policy update.

**Actor Network Update.**

For the subsystem $k$ at timestep $t$, the parameters of the actor network determine the policy $\boldsymbol{\pi}$, which gives the acceleration sets $A_k^t$ under a certain state $S_k^t$. Specifically, the update of the actor network is to maximize the objective function, which is defined as the following:

$$L^{CLIP}(\theta) = \hat{E}_t[\min(r_t(\theta)\hat{A}_t, clip(r_t(\theta), 1-\varepsilon, 1+\varepsilon)\hat{A}_t] \quad (20)$$

where $\theta$ is the policy parameter; $r_t(\theta)$ is the probability ratio of the new policy and old policy at time t, which is formulated as $r_t(\theta) = \frac{\pi_\theta(a_t|s_t)}{\pi_{\theta_{old}}(a_t|s_t)}$; Ɛ is the clipping parameter; $\hat{A}_t$ is the estimated advantage of taking action $a_t$ at timestep t. A truncated version of generalized advantage estimation is adopted to calculate $\hat{A}_t$:

$$\hat{A}_t = \delta_t + (\Upsilon\lambda)\delta_{t+1} + \cdots + \cdots + (\Upsilon\lambda)^{T-t+1} \ \delta_{T-1}$$

(21)

where $\Upsilon$ is the discount factor, $\lambda$ is a hyperparameter to adjust the trade-off relationship between bias and variance; $T$ is the sampled minibatch size; $\delta_t = r_t + \Upsilon V_\Phi(s_{t+1}) - V_\Phi(s_t)$

Thus, $r_t(\theta)$ is clipped at $1-\varepsilon$ and $1+\varepsilon$ to avoid a large update that makes a big difference between the old policy and the new policy. This helps improve the converging performance for the DPPO agent to explore the huge continuous dimension space of the acceleration sets $A_k^t$. The parameters of policy $\pi_\Theta$ are updated with the gradient $\nabla L^{CLIP}(\theta)$ for the objective function:

$$\theta = \theta - \eta_\theta \nabla L^{CLIP}(\theta) \quad (22)$$

where $\eta_\theta$ is the learning rate for updating the actor network.

**Critic Network Update.**

Critic network provides evaluation for the output action $A_k^t$, which helps maximize the objective function $L^{CLIP}(\theta)$. The goal of updating the critic network is to minimize the critic loss $L_c(\Phi)$, which is defined as follows:

$$L_c(\Phi) = \hat{E}_t(R_t(\lambda) - V_\Phi(s_t))^2 \quad (23)$$

where $V_\Phi(s_t)$ is the estimated value from gradient network, with parameters updated through the gradient $\nabla L_c(\Phi)$; $R_t(\lambda)$ is the target value calculated by a weighted sum of n-step cumulative rewards $R_t$ at time step t:

$$R_t(\lambda) = (1 - \lambda) \sum_{n=1}^{T-t} \lambda^{n-1} R_t^{(n)} \quad (24)$$

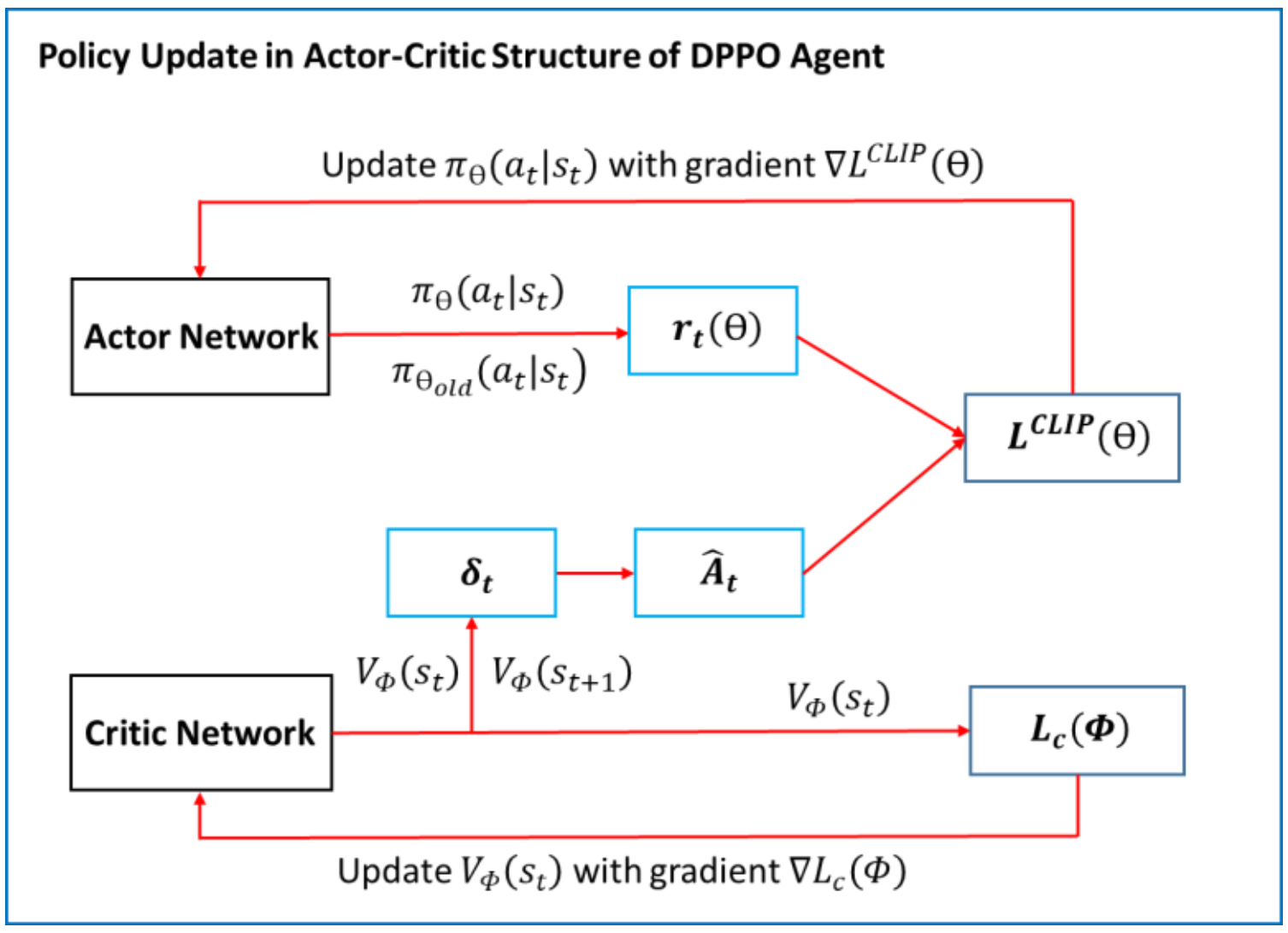

**Fig. 5.** The principle of policy update for DPPO agent

Compared to the PPO algorithm with only one agent, the DPPO algorithm involves one global agent and multiple parallel agents. During the training process, parallel agents interact with environments to collect data for the global agent, while the global agent is responsible for updating the actor network and critic network. The parameters of the DPPO algorithm for the training process are shown in Table 1:

**Table 1**

Parameter Values for DPPO agent

| Parameter | Values |
| --- | --- |
| actor learning rate | 0.00001 |
| critic learning rate | 0.00001 |
| minibatch size $T$ | 512 |
| clipping parameter $\varepsilon$ | 0.2 |
| discount factor $\Upsilon$ | 0.99 |
| number of parallel agents | 4 |

### *3.3. Training procedure and initial settings*

This section discusses the detailed training procedure of proposed RL models. For intuitive understanding, the training process of $\mathrm{M}_4$ with two parallel agents is demonstrated for example. The training environment is built via Python.

Figure 6 shows that the DPPO agent is set to interact with the environment of the control module 4 in real-time. Specifically, parallel agent 1 receives system state $S_{4_1}^{t}$ at each timestep t. Then, agent 1 outputs $A_{4_1}^{t}$ to control CAVs according to the latest policy $\pi(a|s)$ of the global actor network, which moves the current state of the environment forward to the next state $S_{4_1}^{t+1}$. Simultaneously, reward value $R_{4_1}^{t}$ related to $A_{4_1}^{t}$ and $S_{4_1}^{t}$ is computed by the reward function (24), and $S_{4_1}^{t}$ ,$A_{4_1}^{t}$ and $R_{4_1}^{t}$ are stored in the memory buffer. Similarly, agent 2 works in parallel, and the collected data ($S_{4_2}^{t}$, $A_{4_2}^{t}$, $R_{4_2}^{t}$) is sent to the memory buffer every timestep either. The global agent will update the actor network's policy $\pi(a|s)$ and minimize critic loss to update the critic network after a certain batch of data is collected in the memory buffer.

The training process of one episode terminates when the DPPO agent moves to the terminal state $S_4^{\tau}$. Thus, the whole training process from the first state $S_4^{0}$ to the terminal state $S_4^{\tau}$ is defined as an episode. Cumulative rewards of episode $i$, $R_i$ are obtained by adding rewards from timestep 0 to timestep $\tau$. Then, the training process restarts in the next episode and iterates until the last episode is completed.

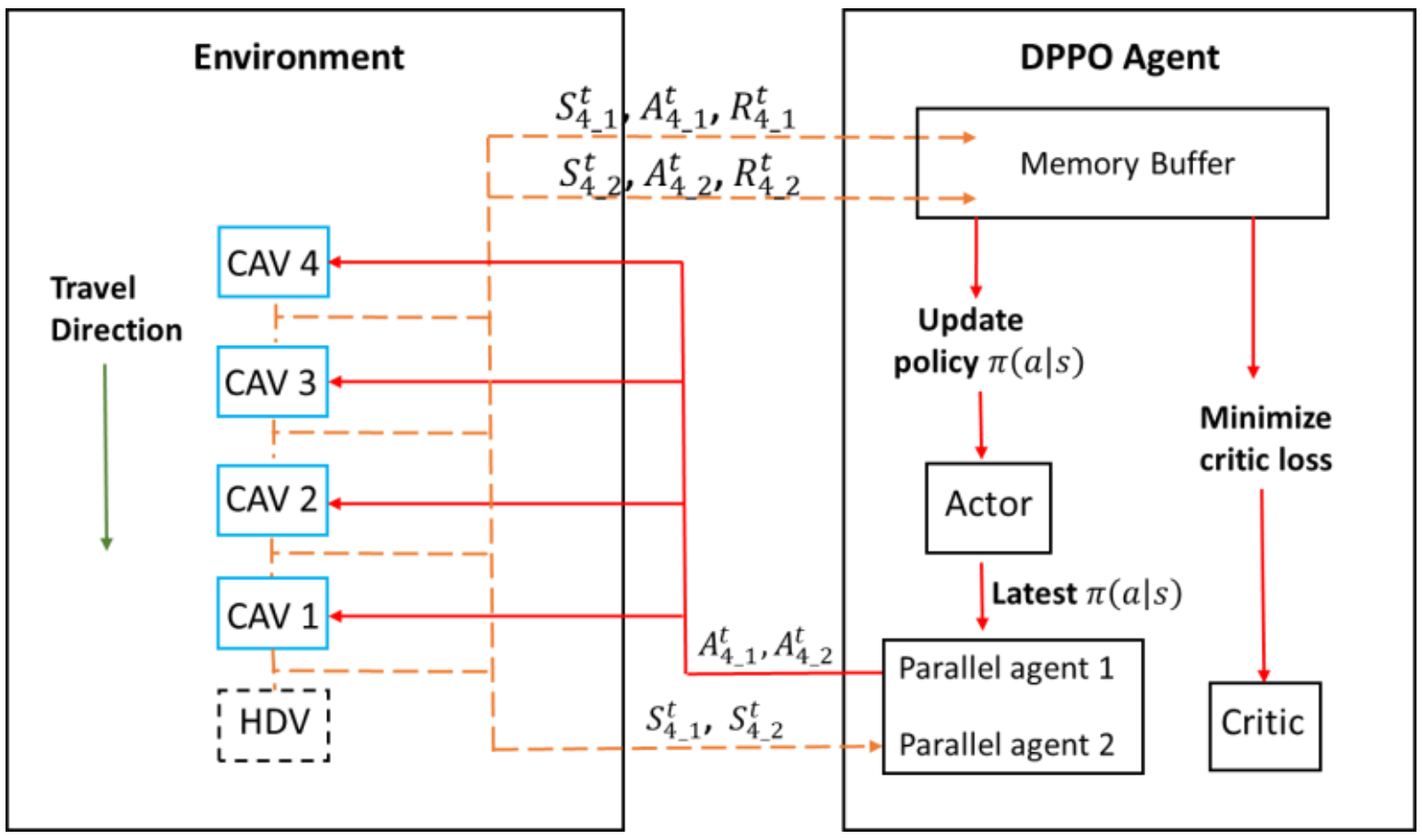

**Fig. 6.** Schematic diagram of training procedure (Environment with two parallel agents)

To incorporate the characteristics of HDV in the control strategy, the trajectory of leading HDV in the training environment is field data from the Next Generation Simulation (NGSIM) datasets, including the typical feature of acceleration, deceleration, and uniform speed, which is shown in Figure 7. The following CAVs starts with equilibrium states, defined as follows:

- Initial speed for CAV $i$ $v_i(0)$: $v_i(0) = v_l(0)$, where $v_l(0)$ is the initial speed of the leading HDV
- Initial spacing distance for CAV $i$ $d_i(0)$: $d_i(0) = v_i(0) * \tau_i^*(0) + l_i$
- Initial gap distance for CAV $i$ $g_i(0)$: $g_i(0) = d_i(0) - l_v$, where $l_v$ is the vehicle length.

The number of training episodes is set to 400 for $\mathrm{M}_1$ and 230 for other models with 218 timesteps for one episode based on the leading HDV trajectory. The updating interval is 0.1s, which coordinates with the sampling frequency of the NGSIM dataset. For $\mathrm{M}_2$, $\mathrm{M}_3$, $\mathrm{M}_4$, and $\mathrm{M}_5$, the continuous action space dimension is enormous, making it difficult for convergence. To address the problem, we trained a model 1 ($\mathrm{M}_1$) that reaches the convergent reward value with insufficient episodes. The unstable acceleration output by the undertrained $\mathrm{M}_1$, shown in Figure 7, is equivalent to acceleration with random disturbances. Thus, in the first 80 episodes of the training process, the rewards are obtained by iterative trajectories generated by undertrained $\mathrm{M}_1$. These are excellent samples that help the DPPO agent explore different dimensions due to the random disturbances and make it easier for the agent to converge to the optimal solution.

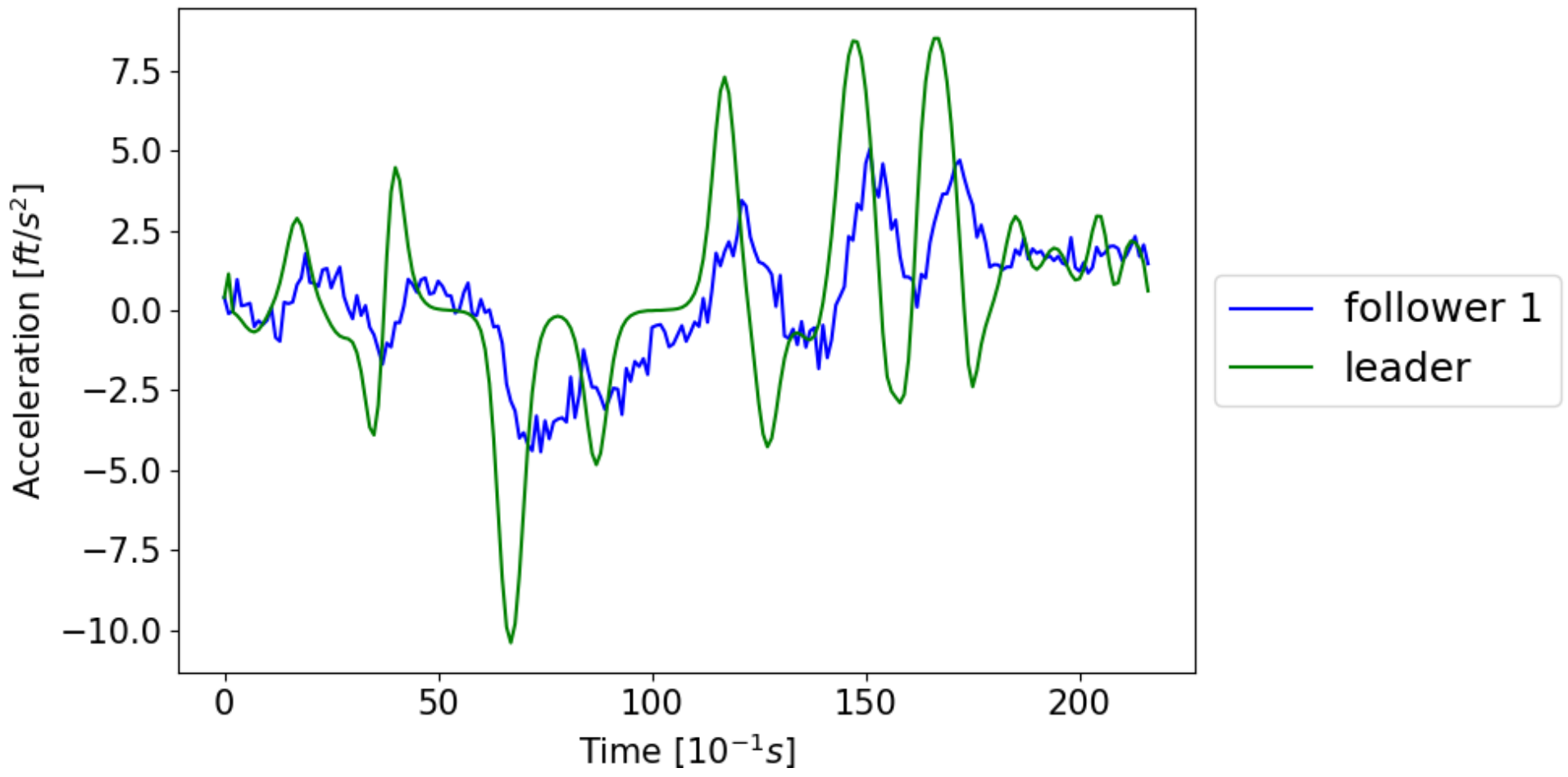

**Fig. 7.** Acceleration trajectory of leading HDV and simulated data by undertrained $\mathrm{M}_1$

In summary, the training process is to obtain the optimal policy, which leads to a control strategy that stabilizes the longitudinal acceleration, maintains the equilibrium state, and reduces energy consumption for CAVs within each module.

## 4. SIMULATION EXPERIMENTS

To validate the performance of the proposed DRL-based models and thus the mixed platoon's control strategy, we conduct simulation experiments via Python embedded with field data from NGSIM datasets of I-80 in California, filtered and reconstructed by (Punzo et al., 2011) and (Montanino & Punzo, 2015). The raw field data was filtered using a low-pass filter to handle the data noises or missingness efficiently. Vehicle trajectories from 4:00 pm to 4:15 pm in Lane 2 are filtered for analysis because there are sufficient samples of traffic oscillations during this period. Particularly, the simulation experiment consists of three parts: (i) model performance analysis and generalization capability validation in the control module, (ii) application of the proposed models in the relative long platoon (pure-CAV platoon and mixed platoon), and (iii) mixed platoon with different combinations of HDV and CAV at the same CAV penetration rate.

Regarding (i), the performance of $\mathrm{M}_1$ and two types of model 5 with different reward functions ($\mathrm{M}_5$ and $\mathrm{M}_5^0$) are analyzed as typical models. The default value settings for the reward function are shown in Table 2, which refers to the parameters (Zhou, et al., 2019). It is worth noting that coefficients $M_i$, $C_i$ and $F_i$ can be adjusted to customize the weights of comfort, car following efficiency, and energy efficiency in reward function. Therefore, different control strategies can be developed through tuning weights. In this paper, we developed two control strategies with different $F_i$ (0/2.5), each of which consists of five models. That means one control strategy ($F_i$ = 2.5) considers the energy efficiency in the reward function, which we mainly analyze. We use $\mathrm{M}_k$ to represent model $k$ for this control strategy. The other strategy does not include energy efficiency in the reward function ($F_i$ = 0), which serves as a comparison. We use $\mathrm{M}_k^0$ to represent model $k$ for this control strategy. Then, the superiority of the cooperative control strategy that $\mathrm{M}_5$ takes is validated in module 5 through comparison with other control strategies (non-cooperative control, 2-2-1 control) based on diverse cooperation degrees. As shown in Figure 8, "non-cooperative control strategy" is to divide the five CAV followers into five individual subsystems, each of which is controlled by $\mathrm{M}_1$. "2-2-1 control strategy", which represents the partially cooperative control strategy between cooperative control and non-cooperative control, is to divide the following five CAVs into three subsystems with CAV follower 1-2, 3-4 and 5 respectively. Both subsystem 1-2 and 3-4 are controlled by $\mathrm{M}_2$ while follower 5 is controlled by $\mathrm{M}_1$. Finally, the generalization capability is validated using large amounts of NGSIM datasets. After validating the proposed models' performance, we implement these models in the experiment (ii) to comprehensively demonstrate its effect of the proposed control strategy on improving traffic flow both in the fully and partially connected automated traffic environment. Regarding (iii), the impact of the combination of CAVs and HDVs in the partially connected and automated environment is studied based on the proposed control strategy. In this way, the effectiveness and robustness of the proposed control strategy are validated comprehensively, and the impact of the combination of CAV and HDV on mixed platoon is analyzed. The default values of parameters, including training settings, vehicle information, and reward function weights, are given in Table 2.

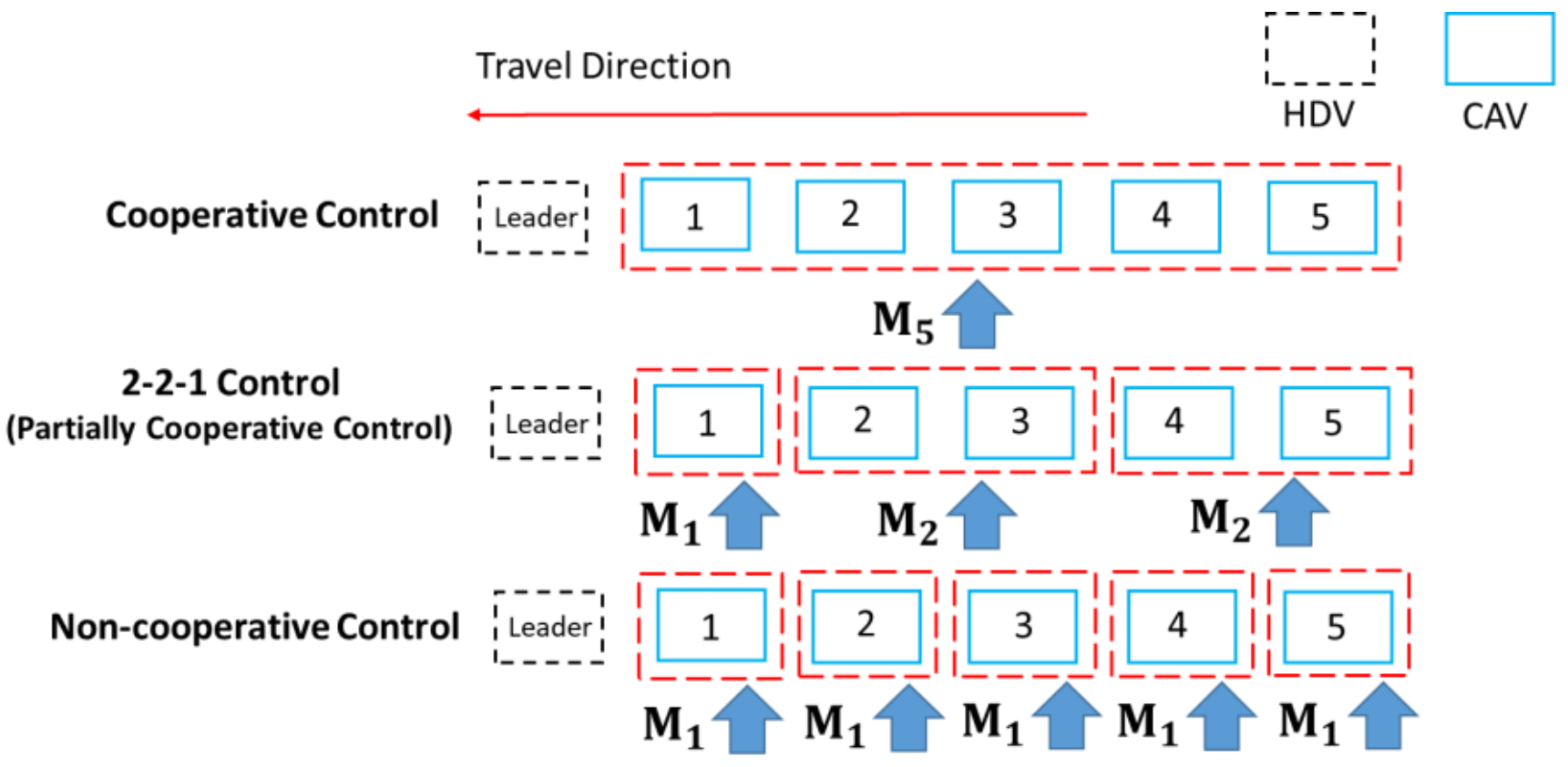

**Fig. 8.** Schematic diagram of different control strategies based on the proposed models

**Table 2**

Default Parameter Setting for the experimental design:

| Parameters | Value |
|---|---|
| Number of subsystem categories $N$ | 5 |
| vehicle length $l_v$ | 15 ft |
| standstill spacing $l_i$ | 21 ft |
| constant time gap $\tau_i^*$ | 1 s |
| $M_i$ | 0.5 |
| $C_i$ | 1 |
| $F_i$ | 0/2.5 |
| $Q_i$ | $\begin{bmatrix} 1 & \\ & 0.5 \end{bmatrix}$ |
| $S_i$ | -0.1 |
| $G_i$ | -0.05 |
| $D_i$ | -0.05 |
| $n_d$ | 50 |
| $g_t$ | 0.6s |
| $v_f$ | 124 ft/s |
| $[a_{i,min}, a_{i,max}]$ | [-13 $ft/s^2$ , 13 $ft/s^2$] |

### 4.1. Training Result

The trajectory of episode reward $R_i$, normally as the indicator for training results, can be noisy due to actor policy's random explorations. Therefore, the moving average reward at episode $i$ $R_i'$ is used to show the training result, which is defined as:

$$R_i' = 0.9R_{i-1} + 0.1R_i \tag{25}$$

The trajectories of moving reward for each model are shown below:

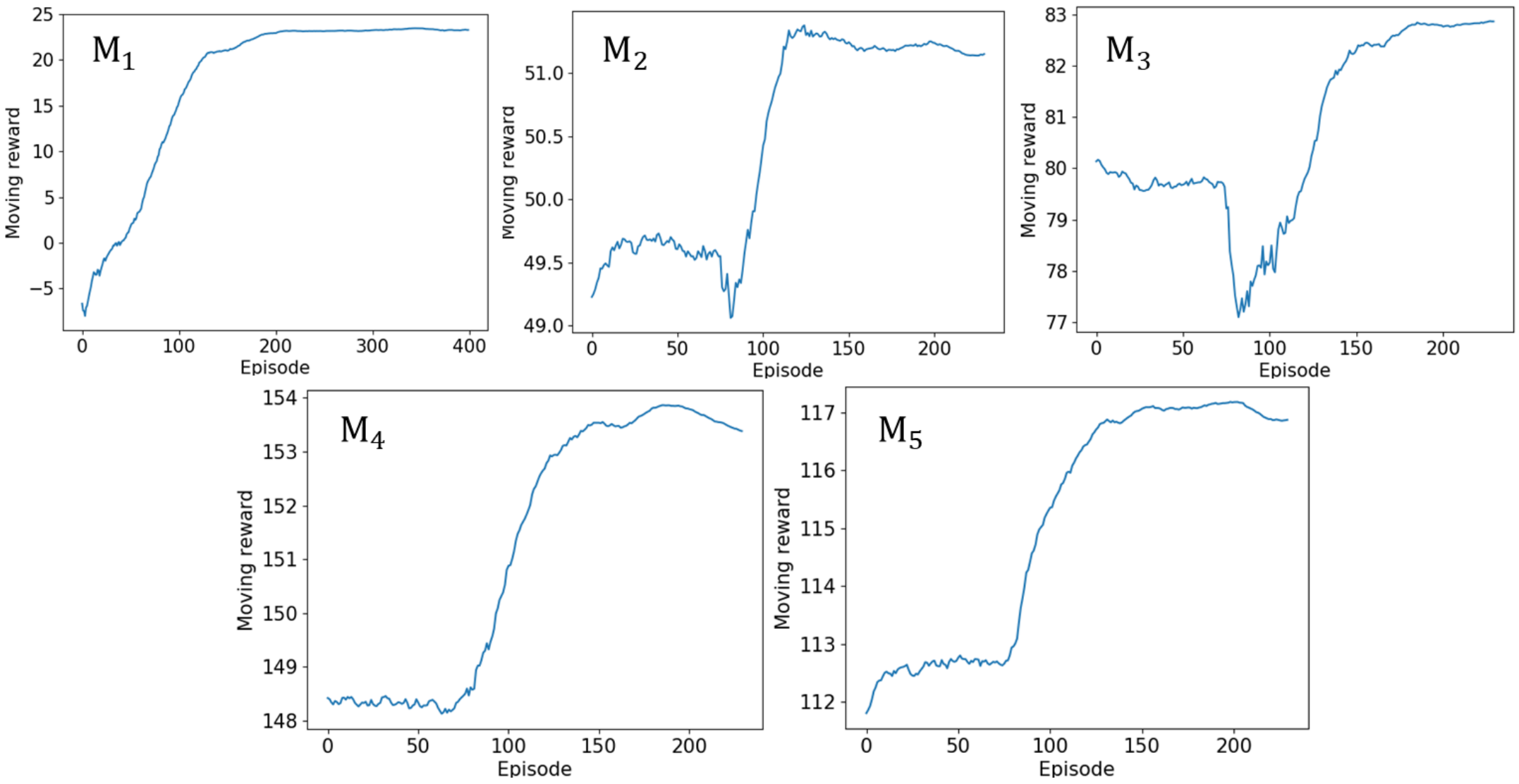

**Fig. 9.** Trajectories of moving average reward for each model

The moving average reward for $M_1$ to $M_5$ shows almost monotonous increases with episodes until stable convergence, suggesting the training procedure's great converging performance. For example, after the first 80 episodes where the guided reward is obtained by untrained $M_1$, the reward value fluctuates in the subsequent several episodes due to dimension exploration and then rises quickly to reach the convergence state.

### 4.2. Model performance validation

This section validates the performance and generalization capability of developed models in the corresponding control modules through five evaluation indexes: cumulative dampening ratio $d_{p,i}$, control efficiency cost $c_i(x_i^t, a_i^t)$, comfort cost $M_i(a_i^t)^2$, running cost $l_i(x_i^t, a_i^t)$ and instantaneous energy consumption $e(x_i^t, a_i^t)$, where the last four indexes are based on the average value per timestep. Field data from NGSIM are compared with simulated data from the proposed models.

### *4.2.1 Performance analysis*

First, we give insight into the performance of the proposed models where $\mathrm{M}_1$ (model 1) and $\mathrm{M}_5/\mathrm{M}_5^0$ (model 5) are analyzed as representatives.

Figure 10 visually shows the trajectories of position, velocity, and acceleration of a two-vehicle platoon of field data and simulated data from proposed $\mathrm{M}_1$. The acceleration for both the leading HDV and the following HDV are fluctuating between -10 $\mathrm{ft/s^2}$ and 10 $\mathrm{ft/s^2}$, which shows poor energy efficiency and driving comfort. A rapid acceleration maneuver of leading HDV at around 38 seconds drastically increases the speed difference and spacing, reflecting the negative impacts of relatively long reaction time by human drivers. In contrast, the CAV follower is more responsive to the leader and maintains a smaller gap with a smoother acceleration within -5 $\mathrm{ft/s^2}$ and 5 $\mathrm{ft/s^2}$. Therefore, the following CAV shows great car following efficiency with little speed difference even when leading HDV accelerates rapidly.

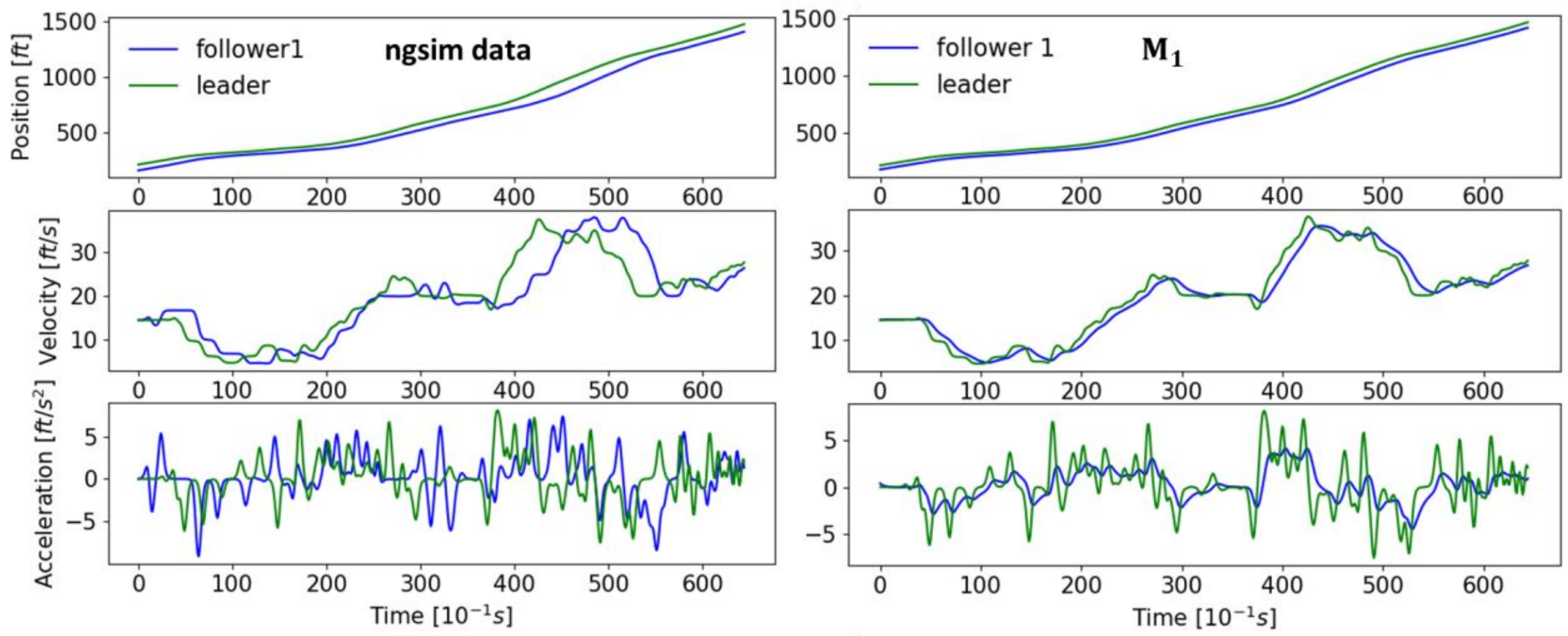

**Fig. 10.** Position, velocity and acceleration trajectories of a two-vehicle platoon

Further, more detailed results are analyzed in terms of the cumulative dampening ratio $d_{p,i}$, driving comfort cost $M_i(a_i^t)^2$, and instantaneous energy consumption $e(x_i^t, a_i^t)$, which draws the same conclusion. The three detailed evaluation indexes of the HDV follower nearly shows no difference with the leading HDV, while a significant reduction of the all three costs (36.3% cumulative dampening ratio, 53.9% driving comfort cost, 11.8% energy consumption) for the CAV follower indicates its excellent string stability, driving comfort, and energy efficiency. Obviously, the smoother acceleration optimizes the three aspects simultaneously because they are all positively correlated with longitudinal acceleration.

Trajectory results for model 5 are shown in Figure 11. In addition to NGSIM data, we also compare the performance of $M_5$ (incorporates energy efficiency in reward function) and $M_5^0$. The trajectory of leading HDV shows frequent stop-and-go waves and standstill periods. Figure 11 (a) shows that the widely varied velocity of leading HDV results in completely stops for following HDVs. Specifically, the increasing stopping time of followers through the platoon indicates amplified disturbance, which intensifies traffic jam. Besides, the irregular spacing of each HDV due to different driving styles further weakened string stability.

Trajectories of simulated data based on different reward functions ($M_5$ and $M_5^0$) show similar results. Compared with the HDV platoon, CAVs show great string stability with non-stopping maneuvers because stop-and-go disturbances are dampened gradually by each CAV. The difference between $M_5$ and $M_5^0$ is that acceleration simulated by $M_5$ is attenuated to a greater extent than $M_5^0$, which indicates $M_5$ that considers energy efficiency in reward function has better dampening performance and thus string stability. This is because there is a coupling relationship between energy efficiency and string stability where energy consumption and dampening ratio are both positively correlated to the acceleration. In other words, the reward aiming to reduce energy consumption by weakening deceleration-acceleration behavior simultaneously improves string stability.

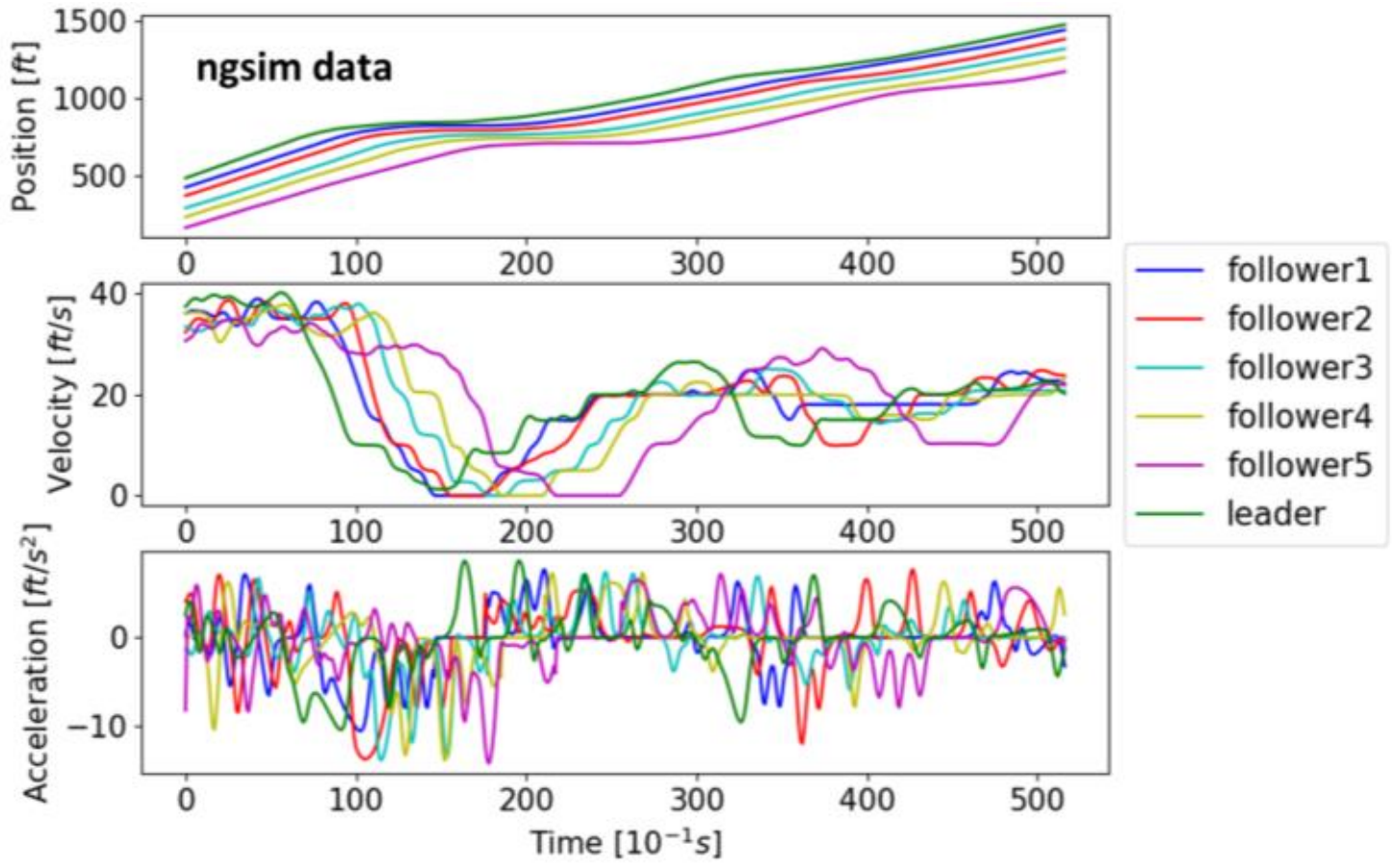

(a) NGSIM data

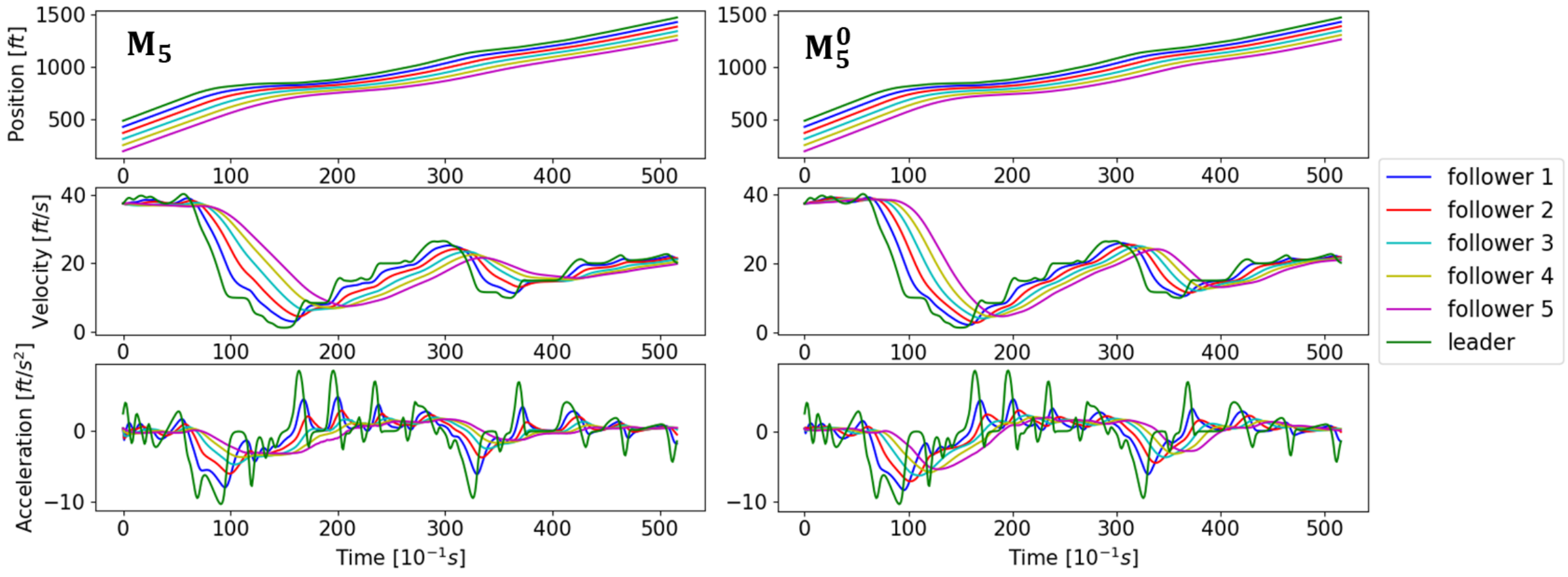

(b) simulated data by model 5 with different reward function weights

**Fig. 11.** Position, velocity and acceleration trajectories of a six-vehicle platoon

Now we give insight to the evolution of control efficiency cost $c_i(x_i^t, a_i^t)$ (evaluation of car following efficiency) of $\mathrm{M}_5$ and $\mathrm{M}_5^0$, which focuses on the deviation from equilibrium spacing $\Delta d_i(t)$ and speed difference with relative leading vehicle $\Delta v_i(t)$. Figure 12 shows detailed results. The diminishing $\Delta v_i(t)$ due to weakened acceleration further validates string stability. On the other hand, the stability of $\Delta d_i(t)$ is sacrificed by taking a cooperative control strategy for both models. Despite unstable $\Delta d_i(t)$, the ranges (-2.5ft to 2ft) for $\mathrm{M}_5^0$ and (-4.5ft to 5ft) for $\mathrm{M}_5$ are tolerable and ensures traffic safety. Compared with $\mathrm{M}_5$, $\mathrm{M}_5^0$ has better performance in control efficiency due to the relatively lower $\Delta d_i(t)$. Thus, there is a trade-off relationship between car following efficiency and dampening performance, which can be adjusted by tuning weights of the reward function.

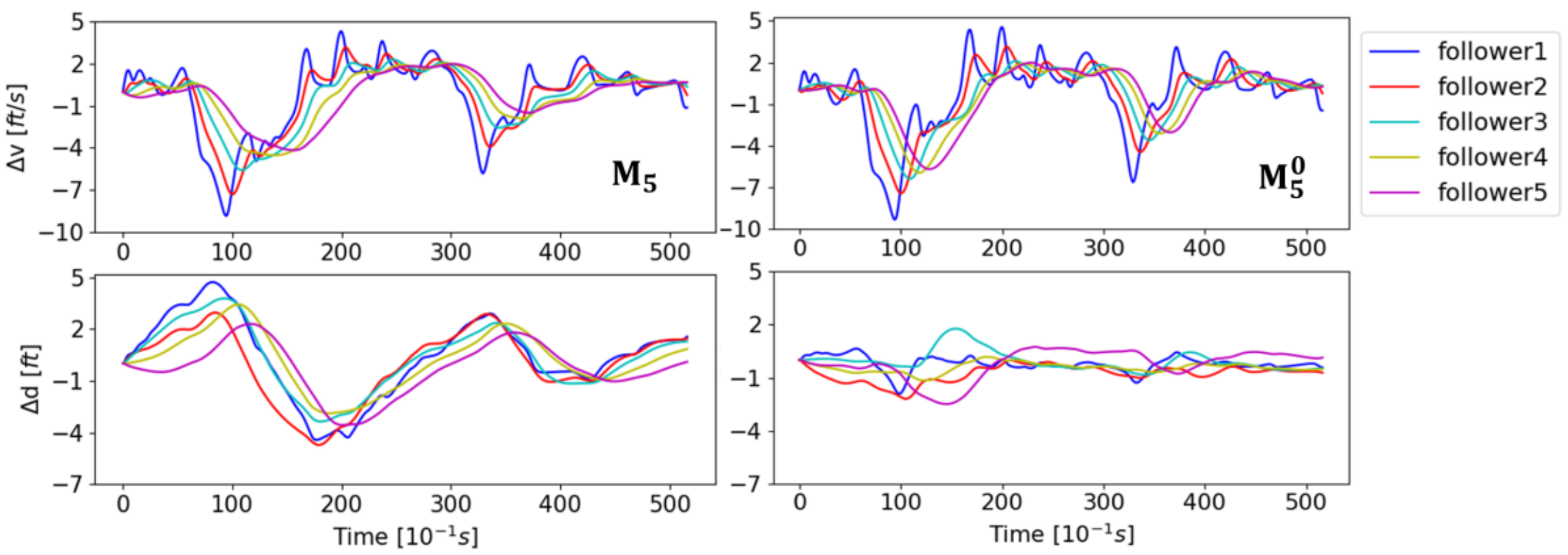

**Fig. 12.** Trajectories of speed difference with relative leading vehicle and deviation from equilibrium spacing

Quantified results are shown in Figure 13. To further validate the superiority of cooperative control by $\mathrm{M}_5$ and $\mathrm{M}_5^0$, we also implemented different control strategies (non-cooperative control, 2-2-1 Control) in the control module 5 based on two sets of models ($\mathrm{M}_k$ and $\mathrm{M}_k^0$). Comparing the results of cooperative control taken by $\mathrm{M}_5$ and $\mathrm{M}_5^0$, we can reach the same conclusion as above. Despite the similar tendency, $\mathrm{M}_5$ shows the lower cost in dampening ratio, comfort cost, and energy consumption while the higher cost in control efficiency cost and running cost compared with $\mathrm{M}_5^0$. Thus, the trade-off relationship between car following efficiency and dampening performance is further validated through the quantified results.

Comparing the results of three control strategies (cooperative, non-cooperative, 2-2-1), the advantage of cooperative control that $\mathrm{M}_5$ takes can be demonstrated, especially in car following efficiency. Three control strategies show minor differences in terms of dampening ratio, comfort cost, and energy consumption. Specifically, the three costs present similar tendencies that drop down through the platoon, demonstrating strict string stability and eco-driving performance. On the other hand, cooperative control outperforms the other two strategies in car following efficiency and local stability due to the downward tendency and the overall lowest control efficiency cost and running cost.

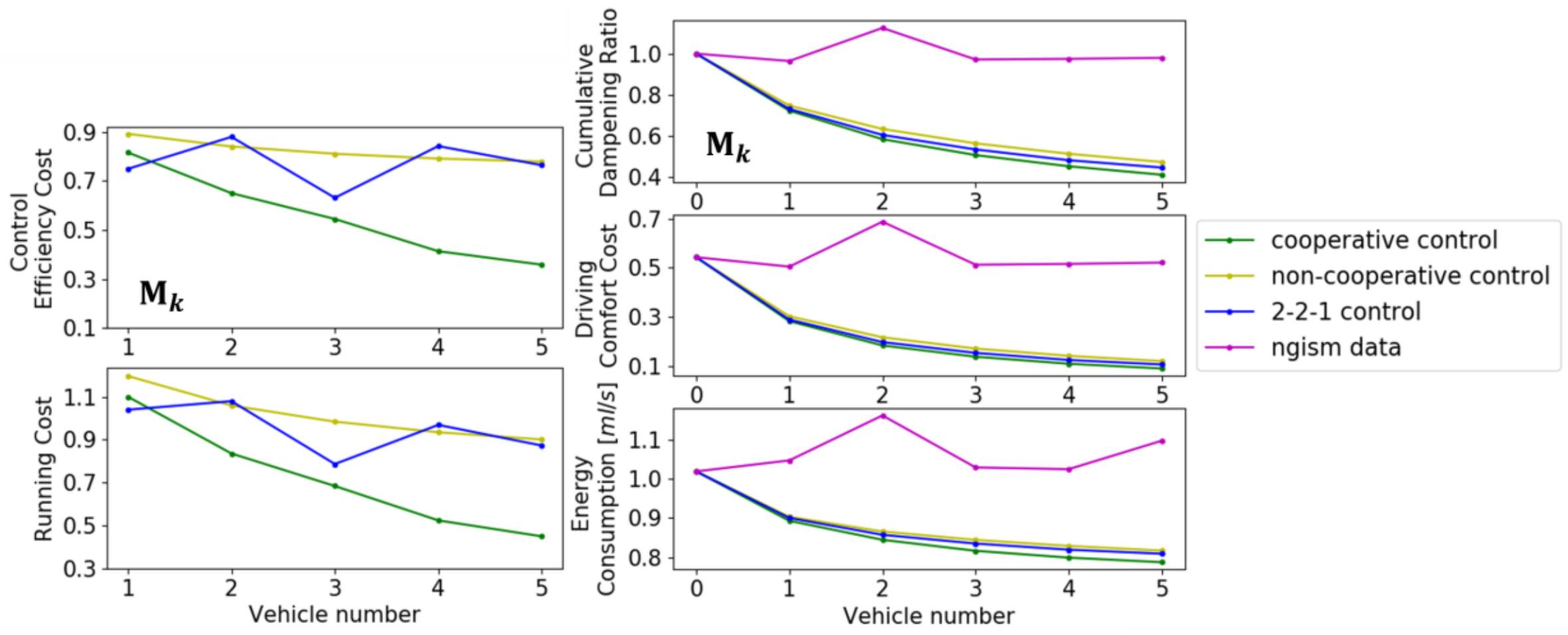

(a) Models ($\mathrm{M}_k$) considering the energy efficiency

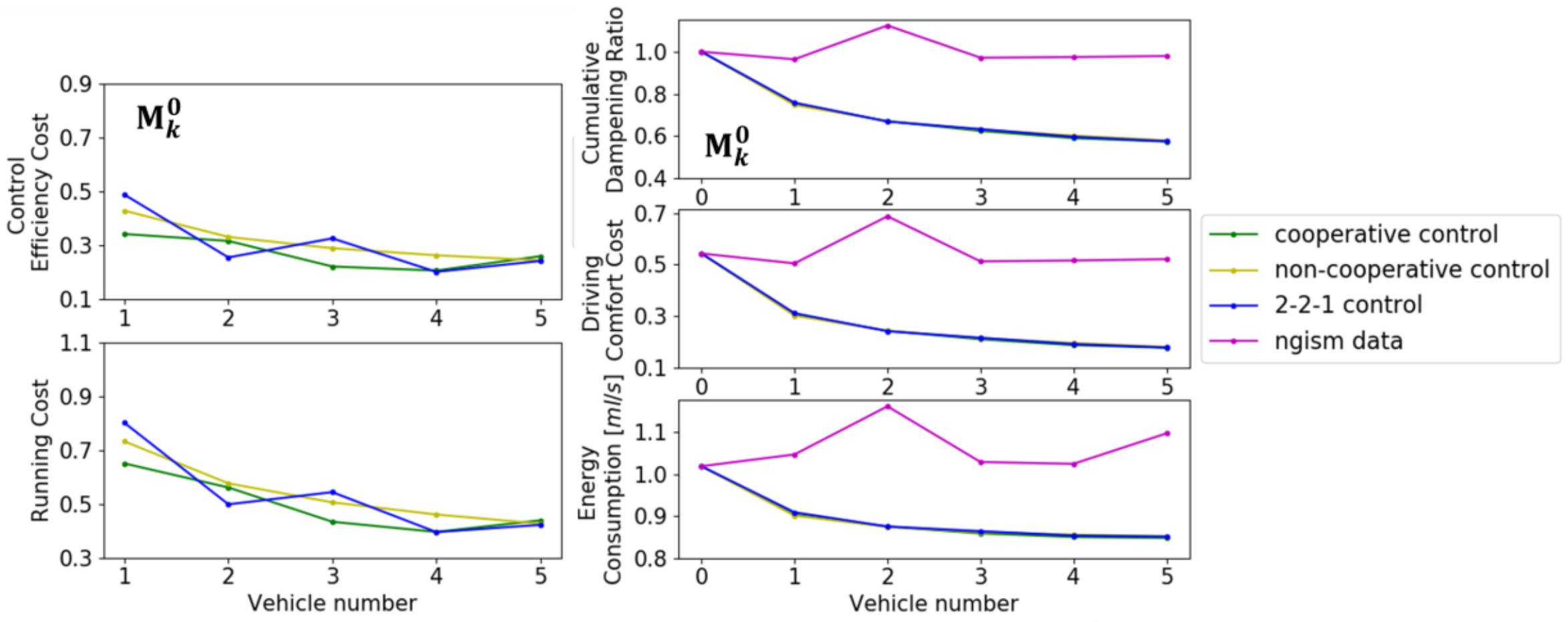

(b) Models ($\mathrm{M}_k^0$) without considering the energy efficiency

**Fig. 13.** Comparison of cooperative control, non-cooperative control, 2-2-1 control and NGSIM data of evaluation indexes for each vehicle (Vehicle number 0 represents the leading HDV)

### *4.2.2 Generalization capability validation*

This section utilizes NGSIM field data excluded from the training process (260 four-vehicle platoons, 224 five-vehicle platoons, and 191 six-vehicle platoons with a time length of over 30 seconds ) as the leading vehicle trajectories to validate the generalization capability of the proposed DRL-based models. We focus on the average performance of the entire control module instead of every single vehicle. The NGSIM datasets are used to compare the generalized performance of cooperative control taken by $\mathrm{M}_5$, $\mathrm{M}_4$, $\mathrm{M}_3$ with other control strategies (non-cooperative control, partially cooperative control (2-2-1 control for $\mathrm{M}_5$, 2-2 control for $\mathrm{M}_4$, 2-1 control for $\mathrm{M}_3$)) and corresponding NGSIM datasets, where the performance of cooperative control taken by $\mathrm{M}_5$, $\mathrm{M}_4$ and $\mathrm{M}_3$ are set as benchmarks. Thus, the percentages of cost reduction by cooperative control compared to other control strategies and NGSIM datasets are shown in Figure 14.

Under the validation of many datasets, the results tend to be consistent in terms of different models. The results show that cooperative control taken by $\mathrm{M}_5$ has greater improvement in performance than $\mathrm{M}_4$ and $\mathrm{M}_3$ due to a higher degree of cooperation, which demonstrates its superiority. Specifically, the cooperative control strategy performs much better than field data in terms of dampening ratio and comfort cost, with significant reductions of at least over 38.96%. Thus, the improvement of string stability and driving comfort is ensured. Besides, energy savings of at least over 14.4% indicates the great eco-driving performance of the proposed models. On the other hand, cooperative control shows great improvement in car following efficiency with at least over 13.63% cost reduction of control efficiency than other strategies.

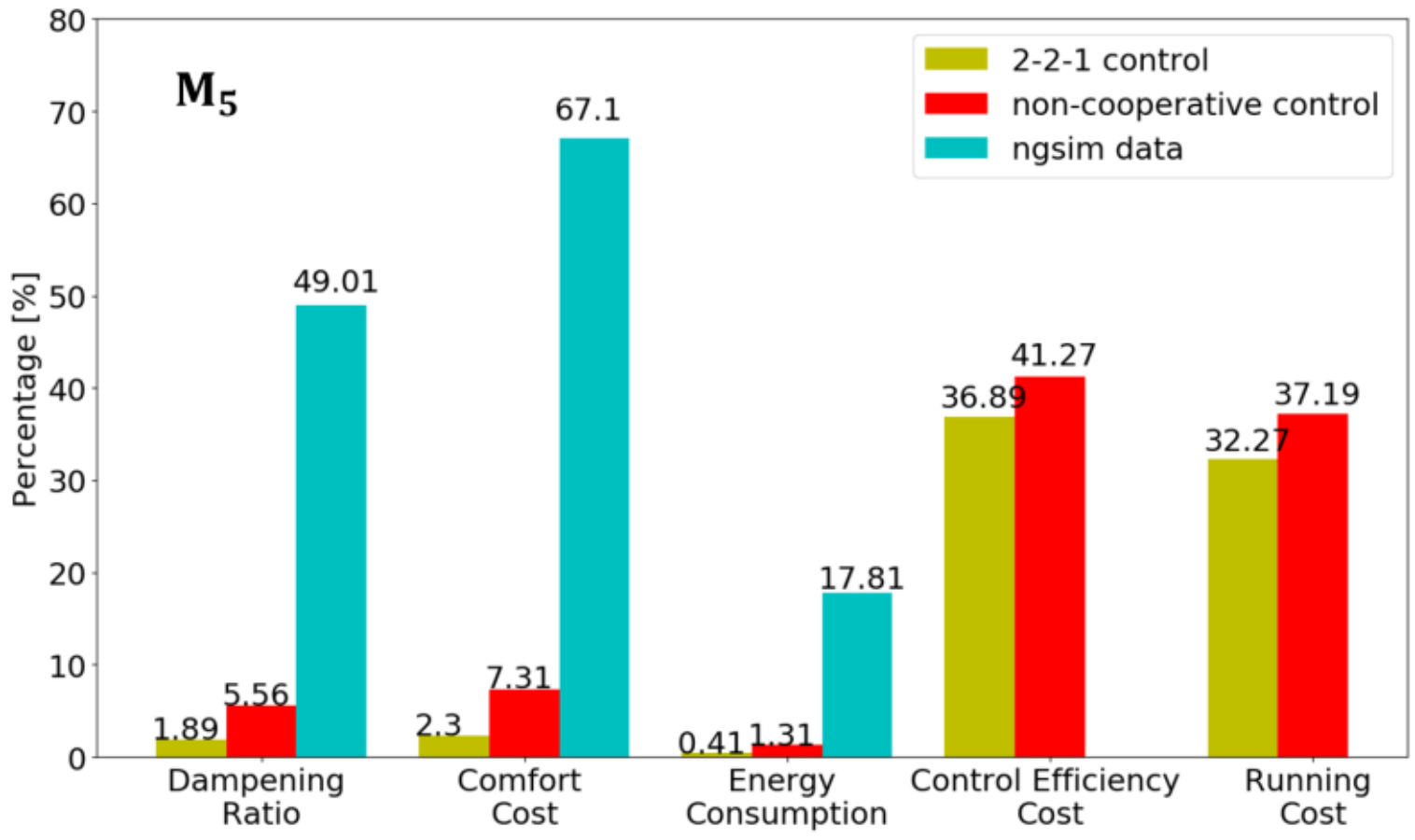

(a) model 5

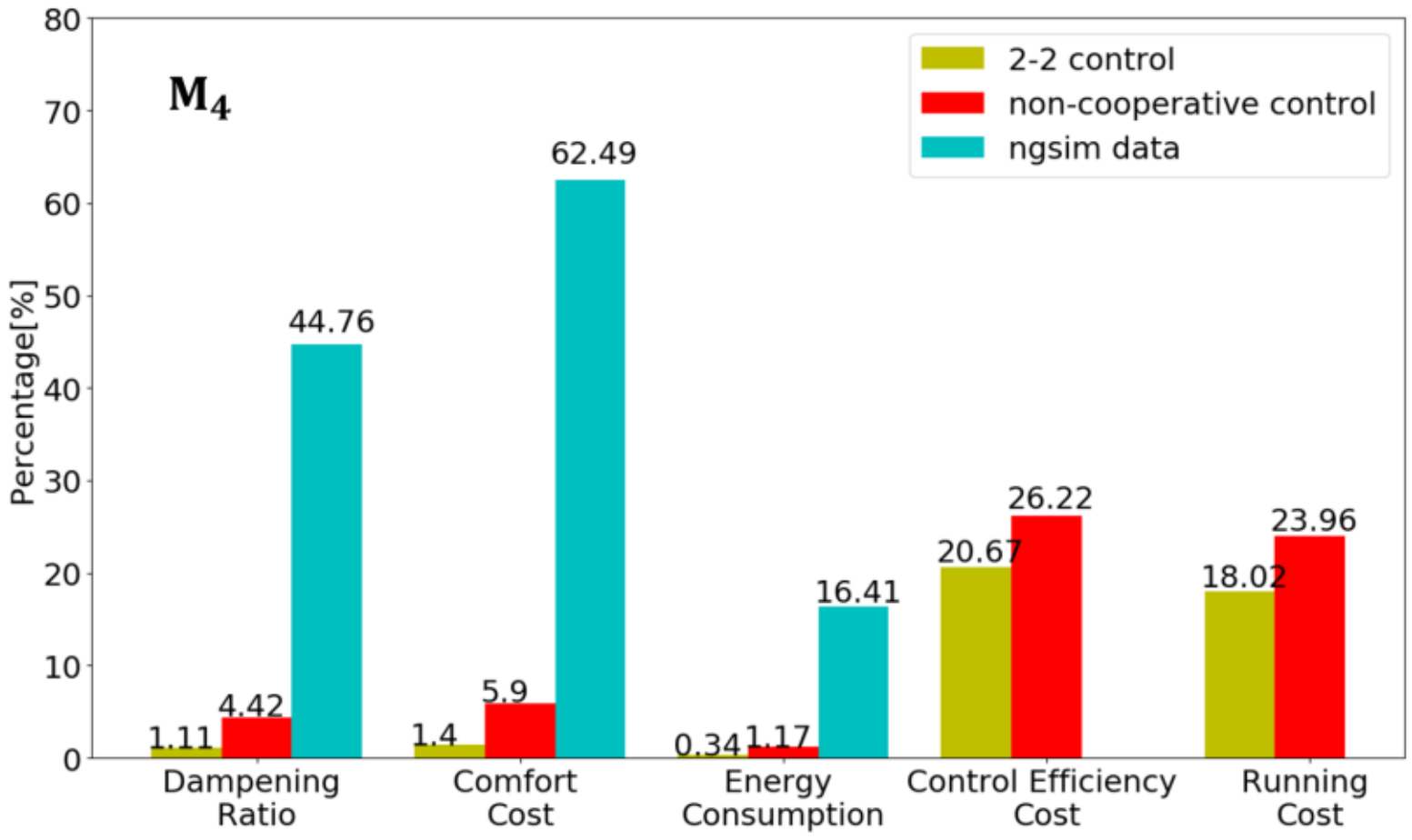

(b) model 4

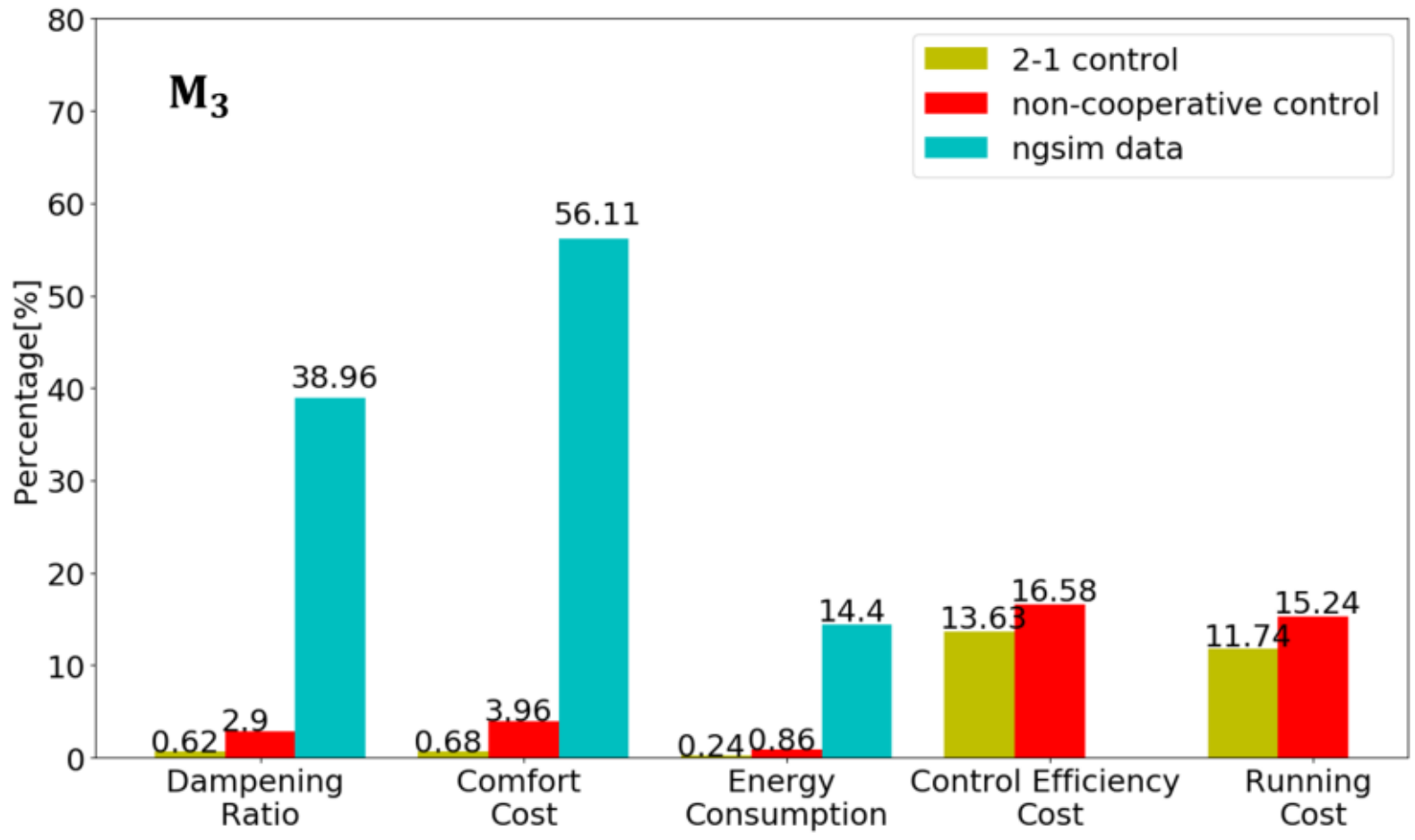

(c) model 3

**Fig. 14.** Performance improvement ratio of cooperative control compared to non-cooperative control, partially cooperative control and NGSIM data through multiple datasets validation (Note: the percentage denotes the ratio of cost reduction compared to other approaches)

To quantify the generalization capability of the proposed models in dampening oscillations and improving energy efficiency, average cumulative dampening ratio and energy consumption are calculated using 333 field trajectories with a time length of over 30 seconds, shown in Figure 15. Each model shows almost the same average values, which suggests great robustness. Particularly, the dampening ratio starts at 1, then monotonically decreases to 0.32 from HDV leader to CAV 5 with average energy consumption falling from 1.23 ml/s to 0.88 ml/s respectively. Thus, the proposed control strategy well achieves strict string stability within the module, which is desirable.

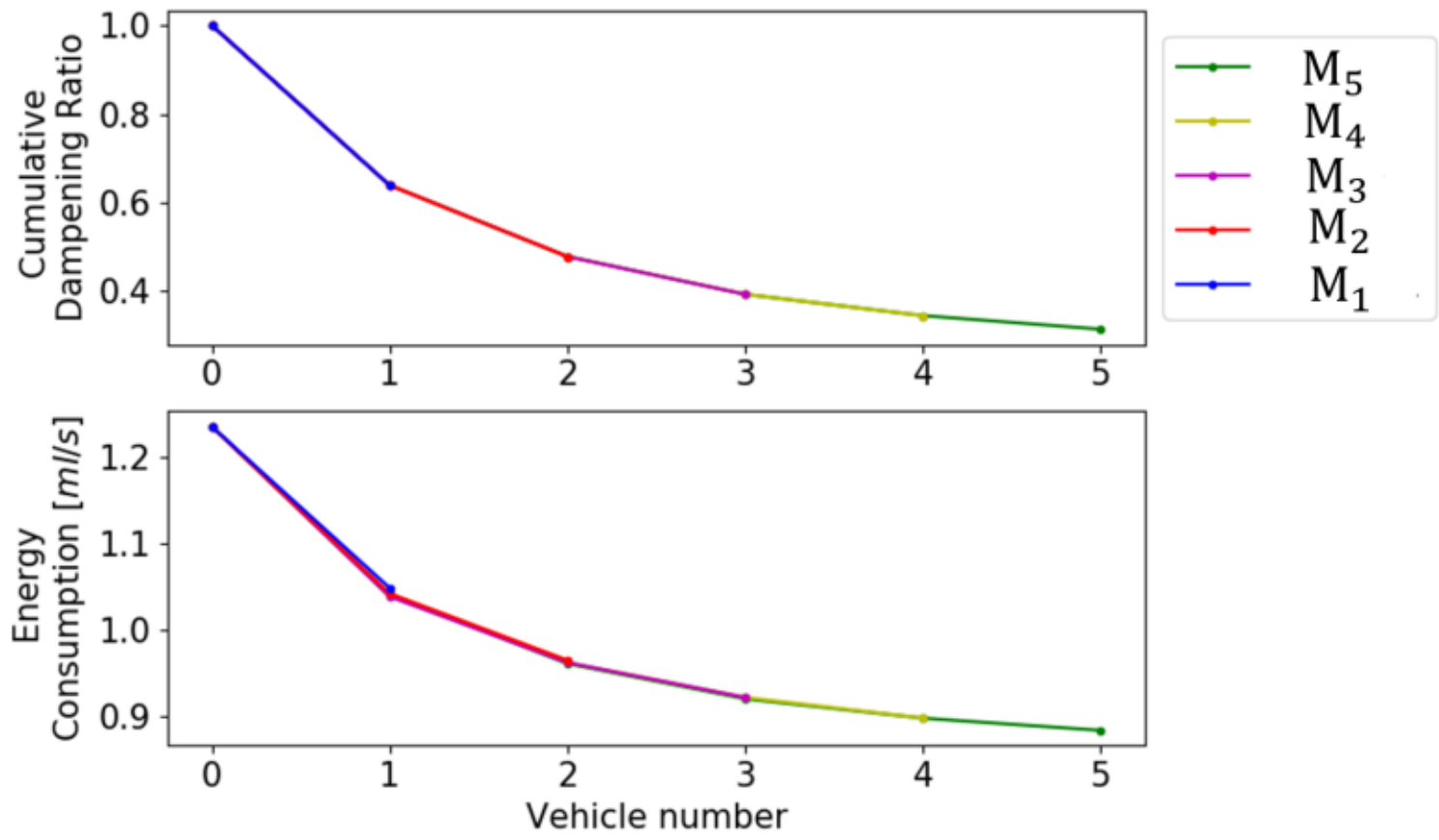

**Fig. 15.** Average energy consumption and cumulative dampening ratio of each model through multiple datasets validation

### *4.3. Application Experiments*

In section 4.3, the proposed models ($\mathrm{M}_k$) control CAVs in both full and partially connected automated traffic environments. For each experiment, the leading vehicle's trajectory is from NGSIM datasets, and 15 followers with initial equilibrium states are configured. Travel efficiency (represented by average velocity), string stability (represented by minimum velocity), and energy efficiency (represented by instantaneous fuel consumption) of the whole platoon are analyzed.

The intelligent driver model (IDM), which has been found to have the best performance to predict trajectories of HDVs among classical car-following models (Zhu, et al., 2018), is adopted to generate trajectories of HDVs. The parameters are calibrated by datasets that include complex situations with several acceleration periods, deceleration periods, and standstill periods (Kesting & Treiber, 2008).

### *4.3.1 Comparison of pure CAV platoon and HDV platoon*

To evaluate the effects of the proposed control strategy on improving traffic flow of fully connected automated traffic environment in terms of travel efficiency, string stability, and energy efficiency, the proposed models ($\mathrm{M}_k$) are applied to control CAVs in the pure-CAV platoon with IDM-based HDV platoon as a comparison. The experiment is conducted with a leading trajectory incorporating frequent acceleration/deceleration disturbances and a short period of standstill. Figure 16 shows the detailed results. In the HDV platoon, the amplified propagative oscillations of leading HDV result in increased standstill periods and a longer acceleration process for followers. Thus, travel efficiency and energy efficiency are negatively impaired with increased traffic congestion. Compared with the HDV platoon, disturbances of the CAV platoon are gradually weakened during propagation through the platoon with no standstill periods.

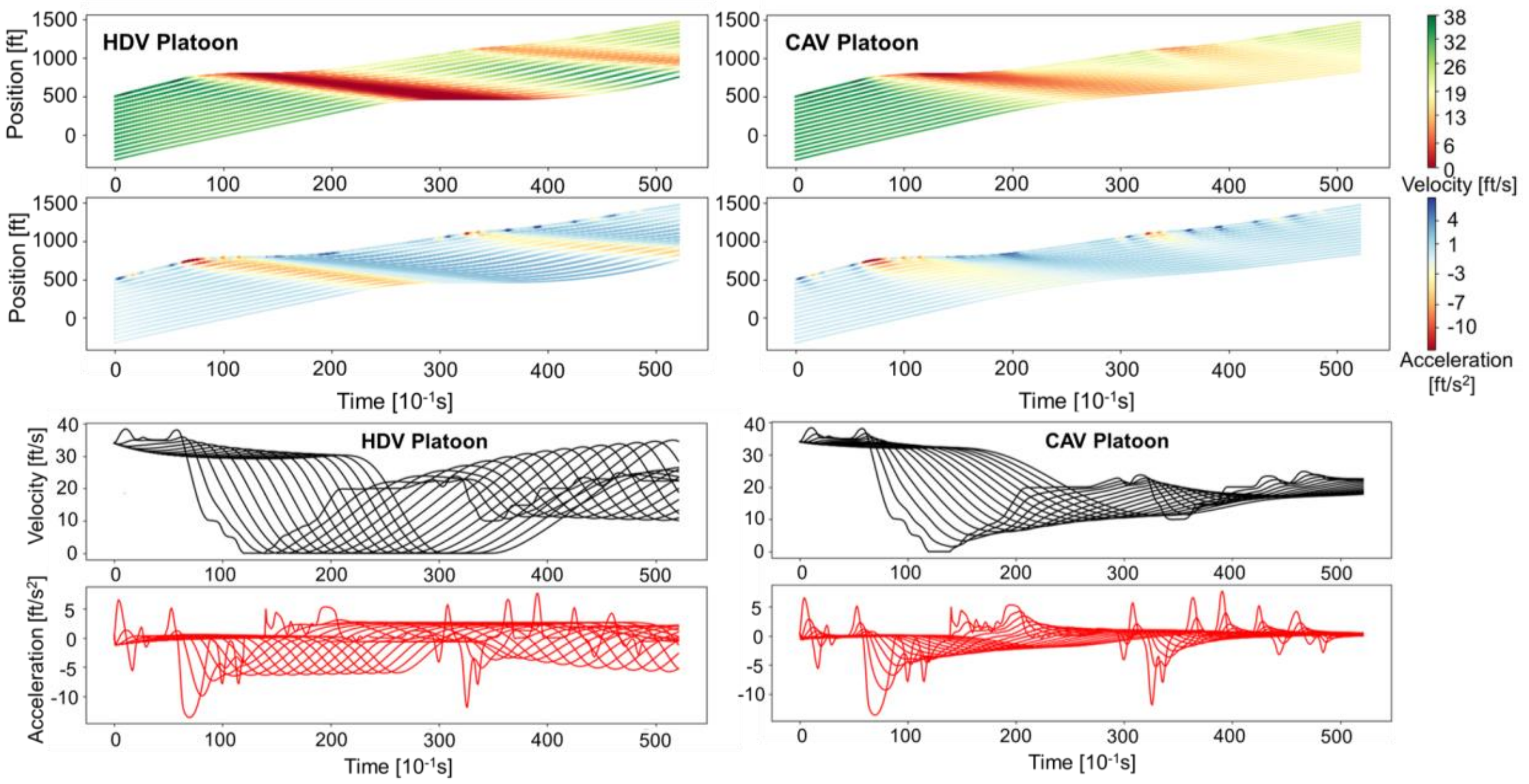

**Fig. 16.** Heat map of velocity and acceleration for HDV platoon and CAV platoon

Three detailed indexes: average velocity, minimum velocity, and average instantaneous energy consumption of each vehicle through the platoon further validate the conclusion, shown in Figure 17. For the CAV platoon, minimum velocity increases monotonically from head to tail due to the improved string stability. Higher average velocity and lower energy consumption through the entire vehicle sequence indicate better travel efficiency and energy efficiency achieved by proposed control strategy.

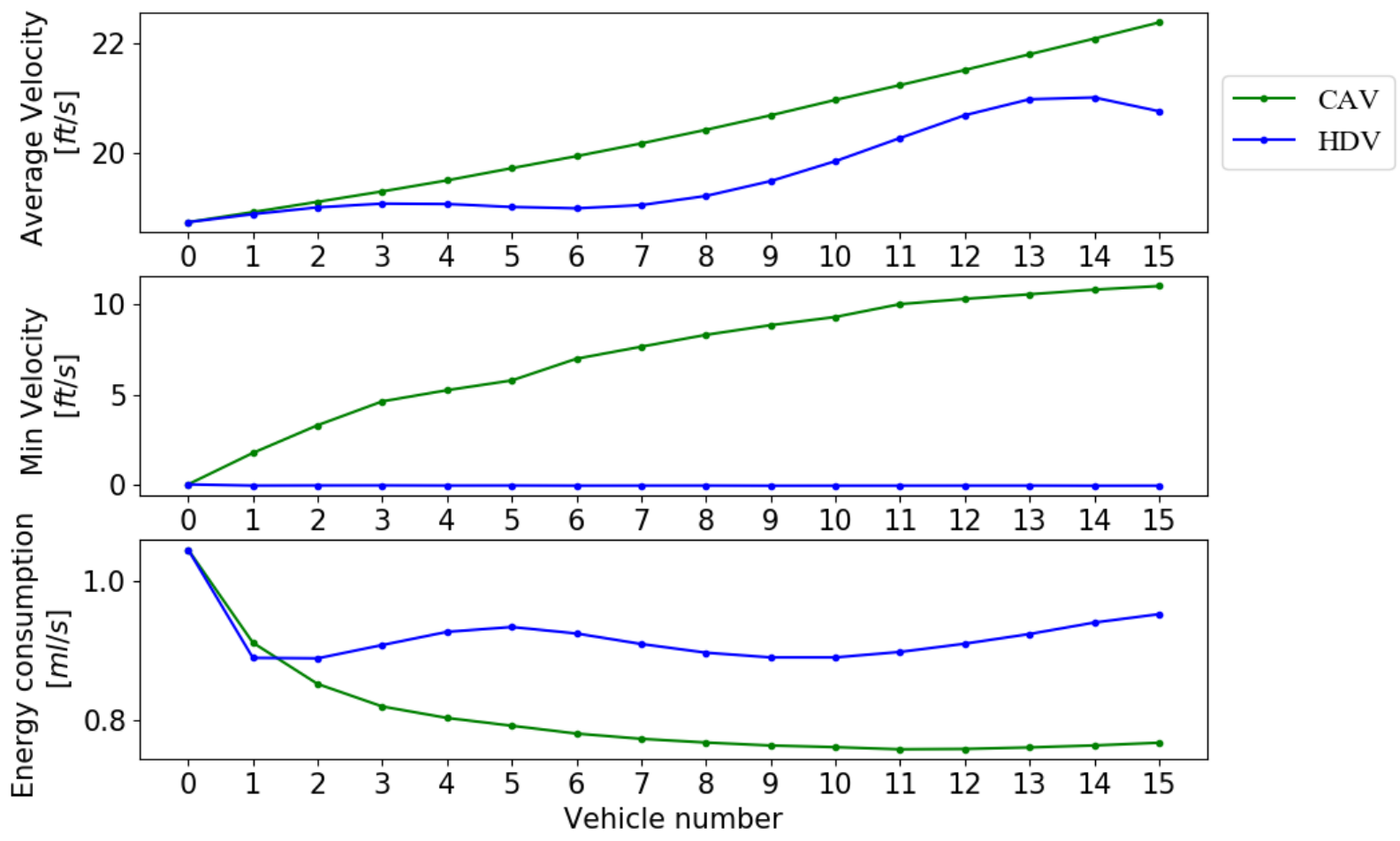

**Fig. 17.** Comparison of detailed indexes of each vehicle in CAV platoon and IDM-based HDV platoon

### *4.3.2 Comparison of mixed platoon with different penetration rates*

To evaluate the effects of the proposed control strategy on improving traffic flow of partially connected automated traffic environment, the proposed models ($\mathrm{M}_k$) are applied to control CAVs in a mixed platoon with six CAV penetration rates (0%, 20%, 40%, 60%, 80%, 100%) where CAVs and HDVs are combined in a random sequence, with detailed results shown in Figure 18 and Figure 19. The trajectory of the leading vehicle has three periods of alternating deceleration-acceleration maneuvers. We can find that dampening performance improves gradually as the penetration rate increases. Specifically, for 0% penetration rate, disturbances amplify towards the end of the platoon in the three periods of acceleration and deceleration, which causes increased standstill duration and longer acceleration process for followers. However, as the CAV penetration rate increases, oscillations' magnitude is mitigating, which gradually weakens the deceleration and acceleration maneuvers. Thus, traffic flow is optimized progressively as the CAV penetration rate increases.

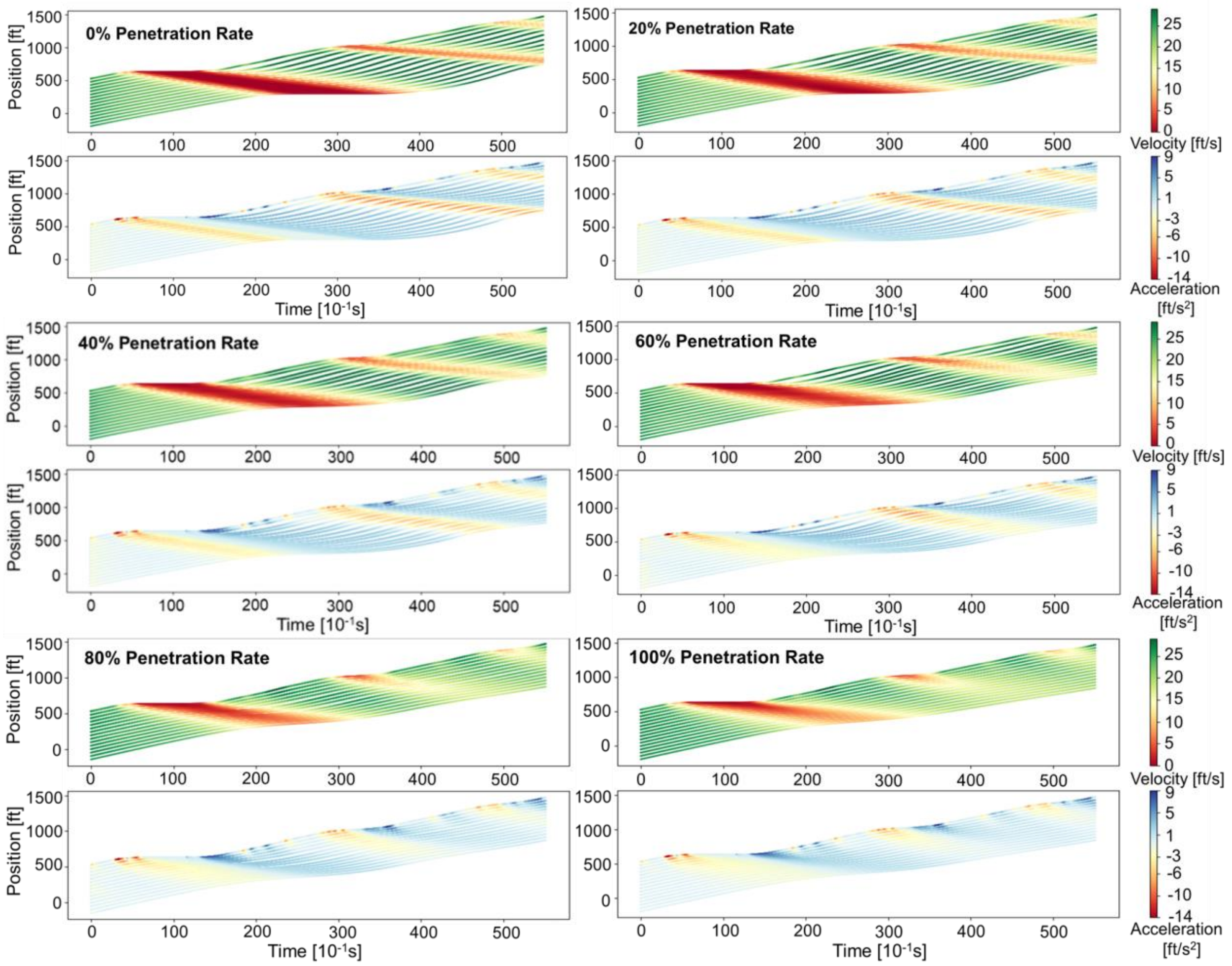

**Fig. 18.** Heat map of velocity and acceleration for mixed platoon with different penetration rates

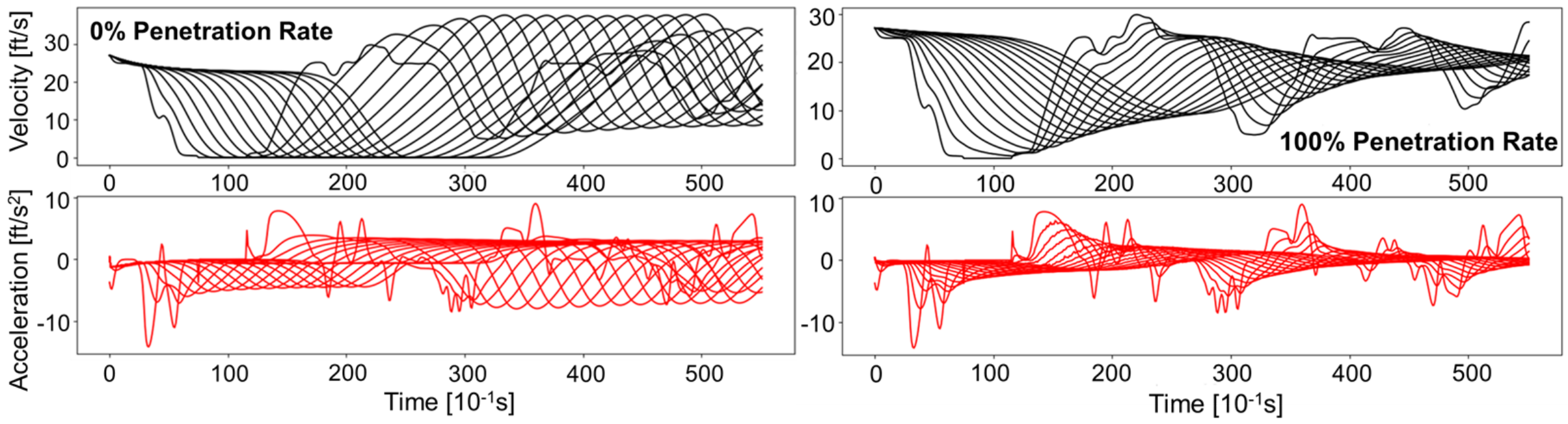

**Fig. 19.** Trajectories of velocity and acceleration for 0% and 100% penetration rates

Multiple datasets (180 field trajectories with a time length of over 50 seconds) are further used to validate the control strategy's generalization capability in the partially connected automated traffic environment and quantify the travel efficiency improvement and energy efficiency improvement in different penetration rates, which is shown in Figure 20. They are both almost linearly increasing with rising penetration rates and 3.87% and 8.14%, respectively, when the platoon consists of all CAVs. Therefore, based on the proposed control strategy, the CAV has the capability to dampen the oscillation and optimize the traffic flow to a great extent.

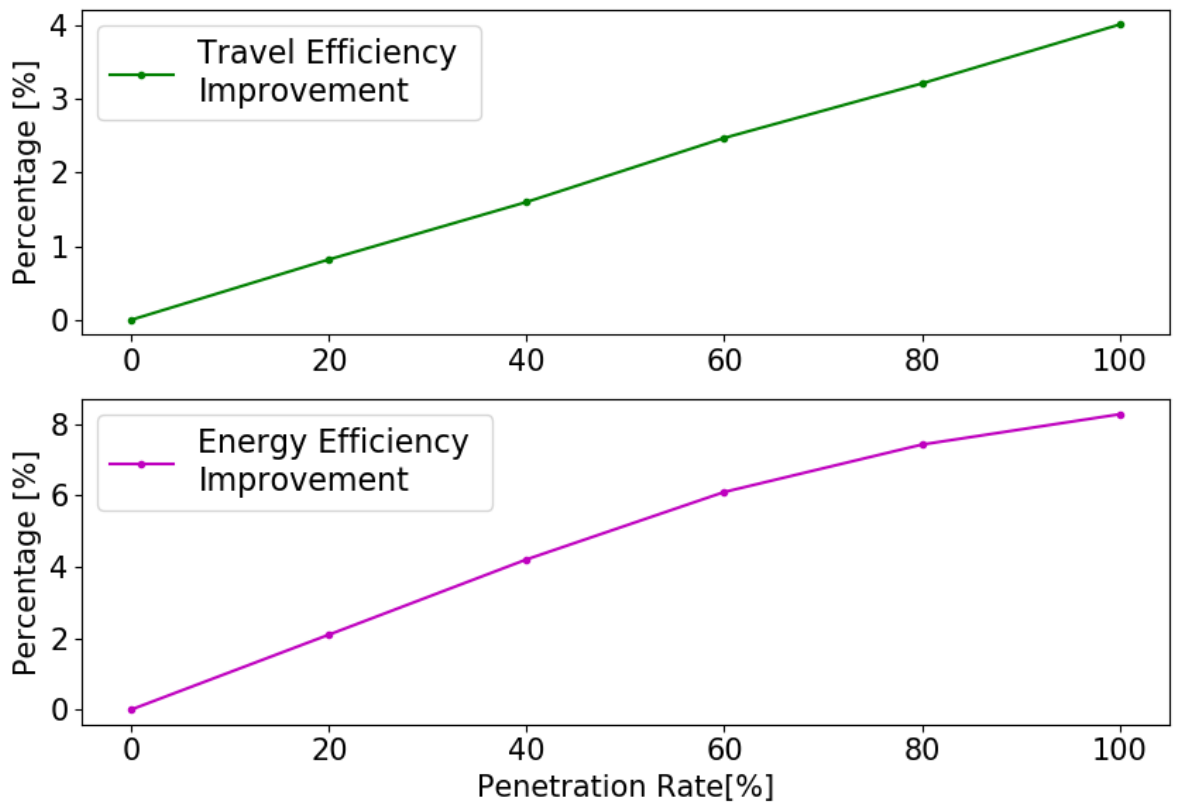

**Fig. 20.** Comparison of average travel efficiency improvement and energy efficiency improvement for the mixed platoon with different penetration rates in low to median scenario (Note: efficiency improvement denotes the percentage of the average speed increment and average instantaneous energy consumption reduction of the whole platoon compared to platoon with 0% penetration rate)

### *4.3.3 Comparison of mixed platoon with different combinations*

Although the proposed control strategy's performance has been validated in the partially connected automated traffic environment, HDVs and CAVs are only combined randomly in previous experiments. However, the combination of CAVs and HDVs can make a significant impact on traffic flow. To evaluate the impact of combinations on mixed traffic flow, four combinations of followers (random combination, specific combination, CAV first, HDV first) are analyzed and compared. 'CAV First' combination and 'HDV First' combination mean all CAVs preceding HDVs and all HDVs preceding CAVs, respectively. Platoon $T_p$ with a specific combination for different penetration rates η is defined as follows:

$$\overrightarrow{T_p} = \begin{cases} (\ 1,1,1,1,0,1,1,1,1,0,1,1,1,1,0\ ), when \quad \eta = \ 20\% \\ (1,0,1,0,1,0,1,0,1,0,1,0,1,0,1), when \ \ \eta = 47\% \\ (1,0,0,0,0,1,0,0,0,0,1,0,0,0,0), when \ \ \eta = 80\% \end{cases} \tag{26}$$

where η represents penetration rate; 1 represents HDV; 0 denotes CAV.

The same configuration in section 4.3.1 is adopted to take the experiment, with results shown in Figure 21. It is cleared that the platoon with all CAVs in front of HDVs outperforms the platoon with other combinations in all aspects (travel efficiency, string stability, energy efficiency) no matter what penetration rate. In contrast, platoon with all HDVs preceding CAVs takes the worst case. Results of "specific combination" and "random combination" show similar performance because HDVs and CAVs are scattered in the mixed platoon for both combinations. The results suggest that clustering leading CAVs can better optimize the entire mixed platoon's traffic flow because oscillations are dampened before they reach HDV followers. Thus, the behaviors of HDVs are optimized to the greatest extent, which mitigates the negative impact of oscillations. In contrast, if HDVs are in front of CAVs, oscillations from the platoon leader are amplified towards upstream, which makes it harder for CAVs to dampen them. Thus, the "HDV first" combination takes the worst case.

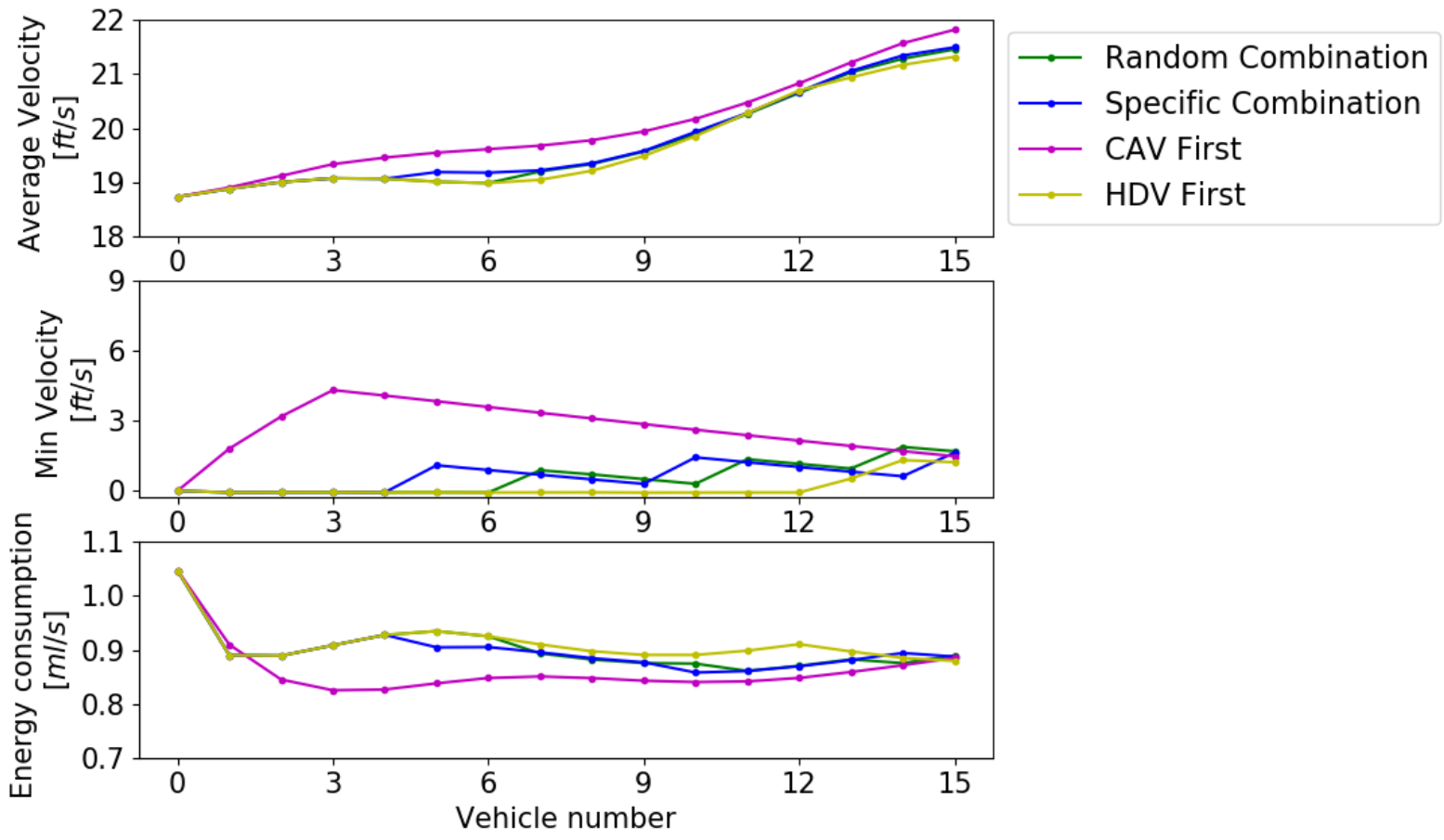

(a) 20% Penetration Rate

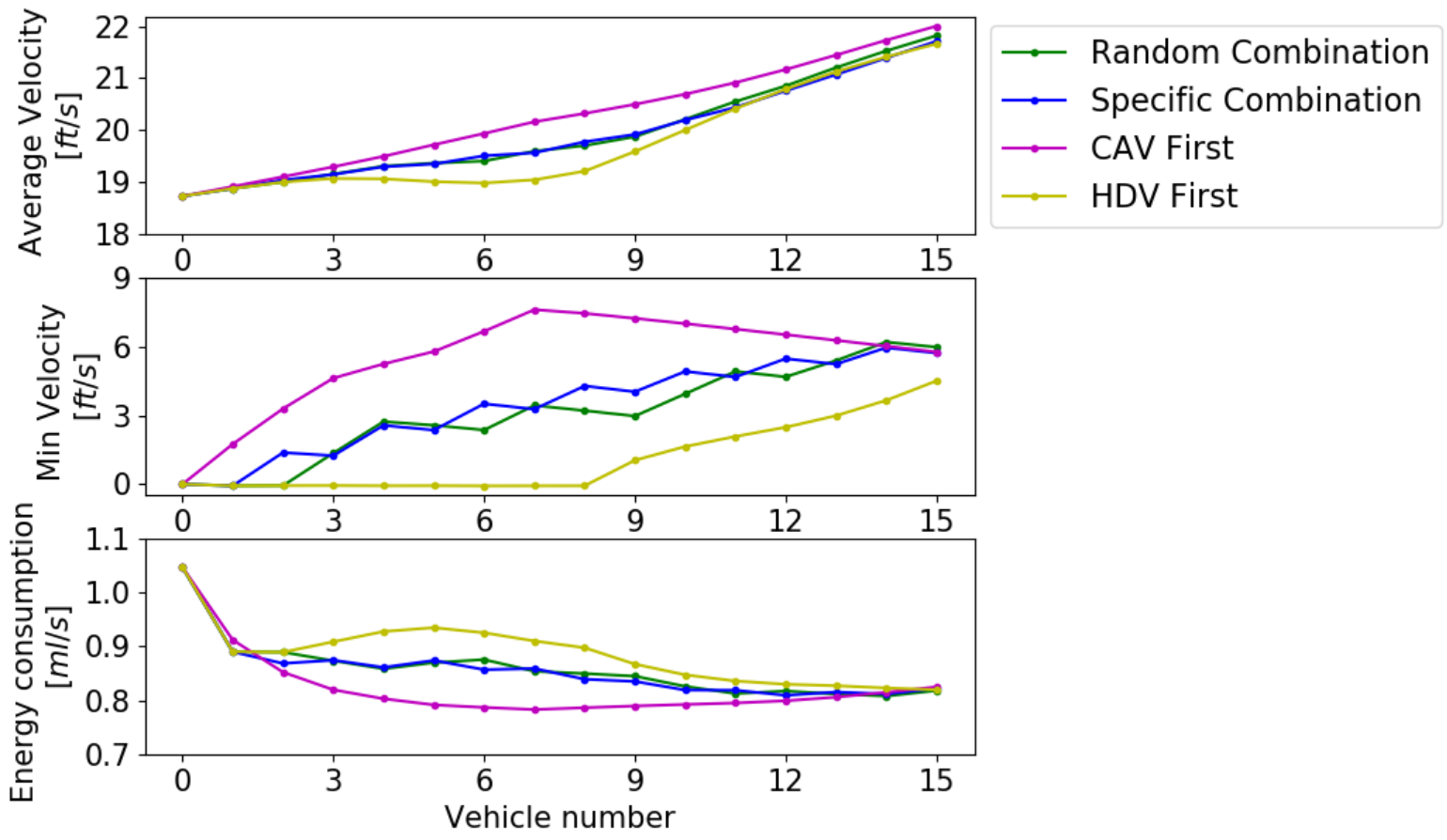

(b) 47% Penetration Rate

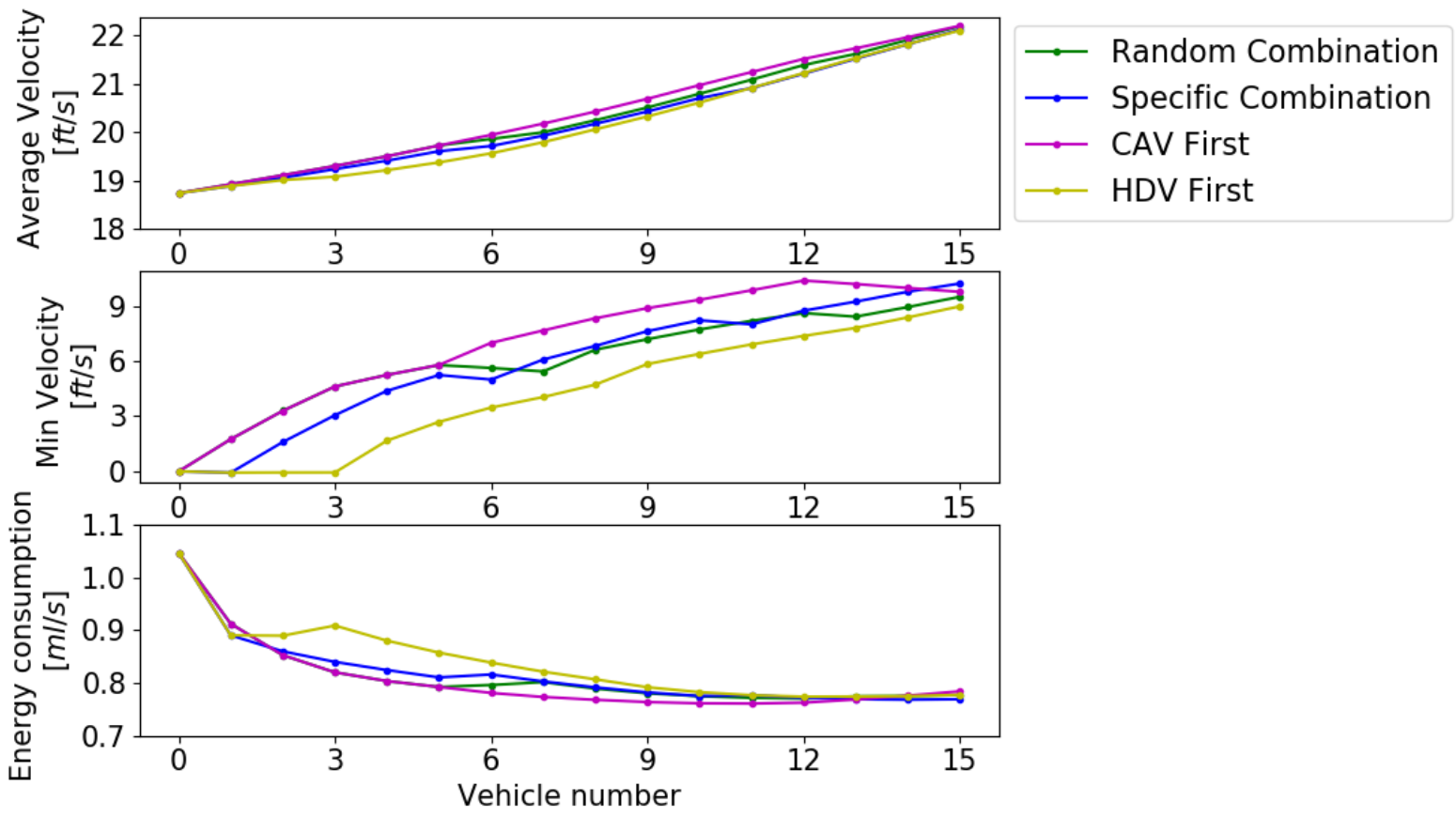

(c) 80% Penetration Rate

**Fig. 21.** Comparison of detailed indexes of each vehicle in mixed platoon with different combinations

The entire platoon's average performance under different combinations further validates the conclusion, shown in Figure 22. There is a great difference between the two extreme combinations ("CAV First" and "HDV First"), while "specific combination" and "random combination" show similar performance. Particularly, the "CAV First" combination improves 2.74% and 7.38% in travel efficiency and energy efficiency compared with the 'HDV First' combination when the CAV penetration rate reaches 47%. Thus, CAVs can be guided to lead the mixed platoon with lane-changing maneuvers, which optimizes traffic flow to the greatest extent.

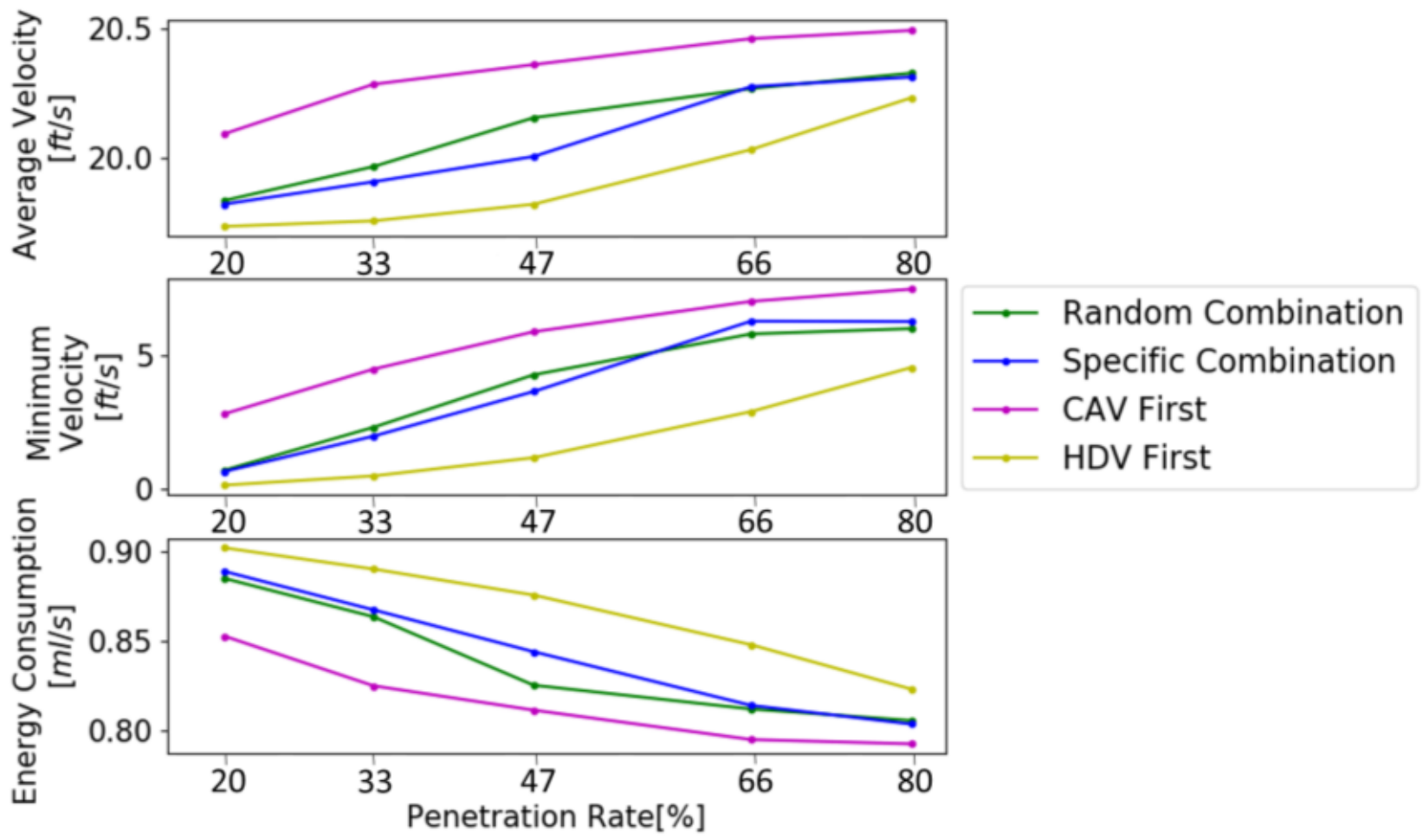

**Fig. 22.** Comparison of detailed results of mixed platoon with different combinations

## 5. Conclusion

This research proposed a cooperative CAV longitudinal control strategy for a partially connected automated environment based on deep reinforcement learning (DRL) algorithm (Distributed Proximal Policy Optimization). The method decomposed mixed traffic into multiple subsystems where each subsystem comprises HDV followed by multiple cooperative CAVs to reduce the training difficulties and dimensions and alleviate communication burdens. Based on that, the longitudinal acceleration of CAVs within each subsystem is cooperatively controlled in the corresponding control modules supported by DRL-based models that we developed. RL agent's reward function was built based on multi-objective function design and equilibrium conditions of car following process to achieve desirable performances of mixed traffic flow optimization in terms of the string stability, car following efficiency, and energy efficiency. HDVs field-collected trajectories were introduced in the training process to incorporate the characteristics of HDVs. Thus, two control strategies based on different reward function weights were developed.

The proposed models were then validated by simulation experiments using NGSIM data. Particularly, the model performance in different aspects, the trade-off relationship between car following efficiency and dampening performance, and the cooperative control strategy's superiority were validated comprehensively. Additionally, the generalization capability of the proposed models was proved through multiple datasets. After that, the proposed models were applied in the simulated partially connected automated environment to demonstrate its desirable performance. Finally, the impact of different combinations of HDVs and CAVs on mixed traffic flow at the same penetration rate was investigated. Results show that platoon with downstream clustered CAVs can achieve the best performance in all aspects because oscillations from leading HDV can be dampened to the greatest extent.

Some future studies can be implemented based on current results. For instance, vehicle dynamics and time-varying communication delay can be incorporated into the control framework to enhance robustness and reliability. Besides, the CAV lateral control for lane-changing maneuvers can be further studied based on the optimized combination of the CAVs and HDVs.

**Acknowledgement**

This work was supported by the Wisconsin Traffic Operations and Safety (TOPS) Laboratory.

**Author Statement**

**Haotian Shi:** Conceptualization, Methodology, Experiment, Analysis, Writing-Original draft preparation. **Yangzhou**: Conceptualization, Methodology, Writing-Review and Editing, Supervision. **Keshu Wu:** Experiment, Writing-Review and Editing. **Xin Wang:** Methodology, Writing-Review and Editing. **Yangxin Lin:** Coding, Writing-Review and Editing. **Bin Ran:** Conceptualization, Writing-Review and Editing.